\documentclass[ reprint,twocolumn,floatfix,amsmath,amssymb,aps,prx,showpacs,longbibliography,superscriptaddress]{revtex4-1}
\usepackage{graphicx}
\usepackage{xcolor}
\usepackage{siunitx}
\usepackage{hyperref}
\usepackage[normalem]{ulem}
\usepackage{here}
\usepackage {amsmath}
\usepackage{bm}
\usepackage{color} \usepackage{graphicx} \graphicspath{{Figures/}}
\usepackage{comment} \usepackage{color}
\usepackage[normalem]{ulem}
\usepackage{placeins}

\renewcommand{\epsilon}{\varepsilon}

\newcommand{\tn}[1]{\textnormal{#1}} 
\usepackage{mathrsfs}
\begin{document}

\title{Wetting-coupled phase separation as an energetic mechanism \\ for active bacterial adhesion}

\author{Dixi Yang}
\thanks{These authors contributed equally}
\affiliation{Advanced Materials Thrust, Function Hub, The Hong Kong University of Science and Technology (Guangzhou), Guangzhou 511453, China}

\author{Anheng Wang}
\thanks{These authors contributed equally}
\affiliation{Institute of Chinese Medical Sciences \& State Key Laboratory of Mechanism and Quality of Chinese Medicine, University of Macau, Macau SAR 999078, China}
\affiliation{Department of Pharmaceutical Sciences, Faculty of Health Sciences, University of Macau, Macau SAR 999078, China}
\affiliation{Zhuhai UM Science and Technology Research Institute, University of Macau, Hengqin 519031, Guangdong, China}

\author{Jia Huang}
\affiliation{Institute of Chinese Medical Sciences \& State Key Laboratory of Mechanism and Quality of Chinese Medicine, University of Macau, Macau SAR 999078, China}
\affiliation{Department of Pharmaceutical Sciences, Faculty of Health Sciences, University of Macau, Macau SAR 999078, China}
\affiliation{Zhuhai UM Science and Technology Research Institute, University of Macau, Hengqin 519031, Guangdong, China}

\author{Xiaofeng Zhuo}
\affiliation{Platform for Structure and Functional Technology of Neural Network, Guangdong Institute of Intelligence Science and Technology, Hengqin, Zhuhai 519031, Guangdong, China}

\author{Chunming Wang}
\email{cmwang@um.edu.mo}
\affiliation{Institute of Chinese Medical Sciences \& State Key Laboratory of Mechanism and Quality of Chinese Medicine, University of Macau, Macau SAR 999078, China}
\affiliation{Department of Pharmaceutical Sciences, Faculty of Health Sciences, University of Macau, Macau SAR 999078, China}
\affiliation{Zhuhai UM Science and Technology Research Institute, University of Macau, Hengqin 519031, Guangdong, China}

\author{Hajime Tanaka}
\email{tanaka@iis.u-tokyo.ac.jp}
\affiliation{Research Center for Advanced Science and Technology, The University of Tokyo, 4-6-1 Komaba, Meguro-ku, Tokyo 153-8904, Japan}
\affiliation{Department of Fundamental Engineering, Institute of Industrial Science, The University of Tokyo, 4-6-1 Komaba, Meguro-ku, Tokyo 153-8505, Japan}

\author{Jiaxing Yuan}
\email{jiaxingyuan@hkust-gz.edu.cn}
\affiliation{Advanced Materials Thrust, Function Hub, The Hong Kong University of Science and Technology (Guangzhou), Guangzhou 511453, China}
  
\date{January 9, 2026}

\begin{abstract}
The rapid adhesion of motile bacteria from dilute suspensions poses a fundamental non-equilibrium problem: hydrodynamic interactions bias bacterial motion near surfaces without generating stable confinement, while electrostatic interactions are predominantly repulsive. Here, combining experiments on \textit{Pseudomonas aeruginosa} and \textit{Staphylococcus aureus} in a polyethylene glycol/dextran aqueous two-phase system with large-scale hydrodynamic simulations, we identify wetting-coupled liquid--liquid phase separation (LLPS) as an energetic trapping mechanism for bacterial adhesion. When bacteria partition into a phase that preferentially wets the substrate, interfacial free-energy minimization creates a deep energetic trap that stabilizes adhesion and induces lateral clustering via capillary interactions. Crucially, bacterial motility plays a dual role: at low phase volume fractions, activity enhances transport into the wetting layer and promotes accumulation, whereas at higher phase volumes it suppresses adhesion through the formation of self-spinning droplets that generate hydrodynamic lift opposing interfacial trapping. Our results establish wetting-coupled LLPS as a generic physical route governing interfacial organization in active suspensions. This provides a unified energetic framework for bacterial adhesion in complex fluids, with broad implications for deciphering bacterial-cell interactions and controlling biofilm formation.
\end{abstract}

\maketitle

\section{Introduction}

The adhesion of microorganisms to surfaces underpins diverse biological phenomena, ranging from biofilm-associated fouling of medical implants and industrial materials~\cite{arciola2018implant,matz2005biofilm,flickinger2017biocoatings,oguntomi2025electrogenetic} to the establishment of complex ecological communities~\cite{fletcher1996bacterial,liang2025recovery,schluter2015adhesion}.
The initial accumulation of bacteria at liquid--solid interfaces constitutes the prerequisite step governing subsequent colonization kinetics~\cite{brading1995dynamics}.
Over the past decades, extensive studies of near-wall bacterial dynamics have identified the key roles of active motility, steric and electrostatic interactions, and long-range hydrodynamic interactions (HIs)~\cite{hori2010bacterial,bechinger2016active,shaebani2020computational}, as well as spatial confinement~\cite{wei2025confinement}, swimming modes~\cite{wu2018entrapment,sartori2018wall}, surface geometry~\cite{sipos2015hydrodynamic}, and bio-chemical signals~\cite{vidakovic2023biofilm,vaidya2025bacteria}.
In particular, HIs reorient swimmers near surfaces, leading to hydrodynamic trapping and enhanced near-wall concentration~\cite{berke2008hydrodynamic,wu2018entrapment,wei2025confinement}.

However, a fundamental paradox persists in the early stages of bacterial accumulation and biofilm formation. Conventional descriptions of bacteria--surface interactions are based on the Derjaguin--Landau--Verwey--Overbeek (DLVO) theory~\cite{hermansson1999dlvo,doi2013soft}, which predicts a substantial electrostatic barrier under physiological conditions due to the like-charged nature of bacterial bodies and solid substrates~\cite{hwang2012adhesion,rijnaarts1999dlvo,israelachvili2011intermolecular}. Such repulsion strongly hinders rapid approach and stable adhesion, particularly at low bacterial densities.
While HIs biases bacterial orientation near interfaces, they are fundamentally dynamical and do not generate an energetic binding potential.
Consequently, hydrodynamic trapping alone cannot stabilize long-lived interfacial confinement, and bacteria readily escape into the bulk under dilute conditions~\cite{wu2018entrapment,wei2025confinement}.
Motility-induced phase separation (MIPS) similarly fails to resolve this paradox, as it requires densities far exceeding those relevant to early colonization~\cite{cates2015motility,fily2012athermal,bechinger2016active}.

Taken together, these considerations indicate that existing frameworks based on DLVO interactions, hydrodynamic trapping, or MIPS cannot account for the rapid formation of dense bacterial aggregates from dilute suspensions~\cite{drescher2013biofilm,lee2023rapid}.
At a fundamental level, these mechanisms either generate repulsive free-energy barriers or rely on dynamical biases that do not create stable minima in the free-energy landscape.
This discrepancy points to a missing energetic mechanism capable of converting interfacial proximity into robust confinement.

Natural bacterial environments are chemically complex and rich in macromolecular polymers, in contrast to the homogeneous fluids considered in most previous studies (Fig.~\ref{fig:figure1}(a), left).
Although polymers can modify bulk swimming behavior~\cite{zottl2019enhanced,torres2024enhancement}, interfacial accumulation in polymeric media obeys distinct physical principles.
Indeed, even simple polymer additives can suppress surface accumulation through viscoelastic lift forces~\cite{cao2022reduced}.
Crucially, many biological polymeric environments --- including mucus, soil, and host tissues --- are heterogeneous and can undergo liquid--liquid phase separation (LLPS), as can bacterial secretions themselves~\cite{ma2021structural,seviour2020phase,gupta2024coli,yoo2019cellular,van2018role}.
In such phase-separated fluids, selective partitioning couples bacterial affinity to interfacial thermodynamics.
We propose that if the phase preferred by bacteria also wets the solid surface, interfacial free-energy minimization creates an energetic trap that confines bacteria at the interface and induces lateral aggregation via capillary forces (Fig.~\ref{fig:figure1}(a), middle and right).
While this organization would be dictated by equilibrium wetting thermodynamics, bacterial motility introduces non-equilibrium transport that can modify both accumulation kinetics and steady states.

In this work, by combining experiments on two distinct bacterial species --- {\it Pseudomonas aeruginosa} (\textit{P. aeruginosa}) and {\it Staphylococcus aureus} (\textit{S. aureus}) --- in a model dextran (DEX)/polyethylene glycol (PEG) aqueous two-phase system (ATPS) near glass walls with large-scale hydrodynamic simulations, we demonstrate that wetting-coupled LLPS provides a robust and general physical mechanism for bacterial accumulation at solid interfaces.
This framework unifies wetting thermodynamics and active transport, overcoming the traditional limitations of homogeneous fluids and moving beyond the classical DLVO theory-based descriptions~\cite{hermansson1999dlvo}.

\section{Results and Discussion}

\subsection{Wettability of surface and bacteria}

\begin{figure}[t!]
\centering \includegraphics[width=8cm]{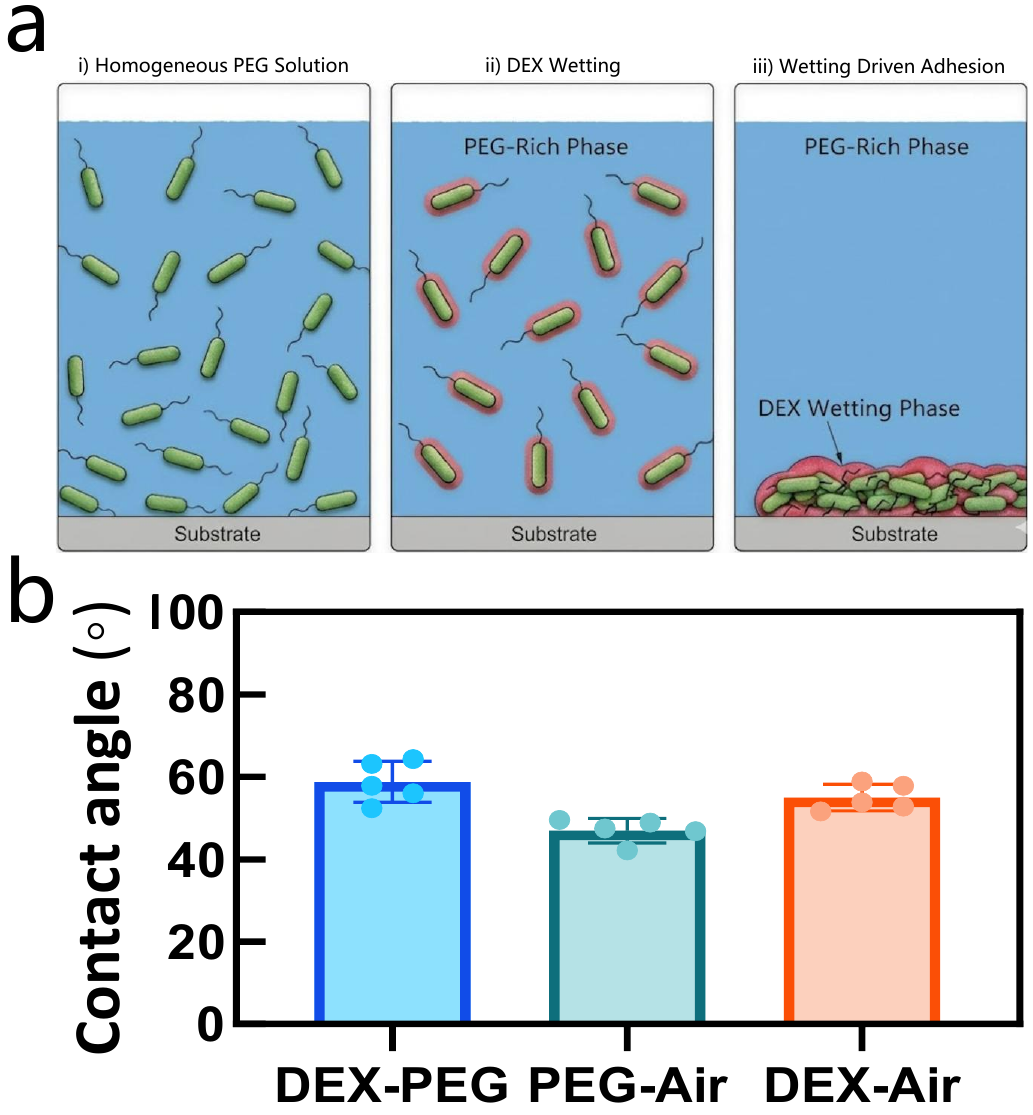}
\caption{Wetting-mediated interfacial adhesion induced by liquid--liquid phase separation. 
(a) Conceptual illustration of the adhesion mechanism. 
(i) In a homogeneous PEG solution, bacterial accumulation at the wall is weak due to the absence of an energetic trapping mechanism. 
(ii) Upon the introduction of a minority DEX phase, bacteria are preferentially wetted by the DEX-rich phase. 
(iii) When the DEX-rich phase also wets the solid surface, a thin wetting layer forms at the interface, energetically confining bacteria near the wall and thereby enhancing interfacial accumulation and in-plane clustering. 
(b) Contact angle measurements of the polymer phases on the glass substrate, demonstrating the preferential wetting of the DEX-rich phase relative to the PEG-rich phase.}

\label{fig:figure1}
\end{figure}

To quantify the wetting properties of the glass walls, we measured contact angles by depositing a DEX droplet onto the PEG--glass interface (see Appendix~\ref{sec:exp} for experimental details). The resulting contact angle $\theta\approx 60^\circ$ indicates partial wetting and a finite affinity of the glass surface for the DEX-rich phase (Fig.~\ref{fig:figure1}(b)). Control measurements on the air--glass interface show a higher affinity of glass for PEG than for DEX (Fig.~\ref{fig:figure1}(b)), emphasizing that the wetting behavior within the aqueous two-phase system (ATPS), rather than in air, is the relevant metric for interfacial organization under experimental conditions. Analogous contact angle measurements performed on bacterial lawns show that \textit{P. aeruginosa} is amphiphilic~\cite{yang2025active}, with a slight preference for the DEX-rich phase (Fig.~\ref{fig:figureS1}(a)). Together, these measurements establish that both the solid surface and bacterial bodies preferentially interact with the DEX-rich phase, satisfying the thermodynamic conditions required for the formation of a surface-associated wetting layer.

\subsection{Phase-dependent bacterial accumulation}

\begin{figure}[t!]
\centering \includegraphics[width=8.5cm]{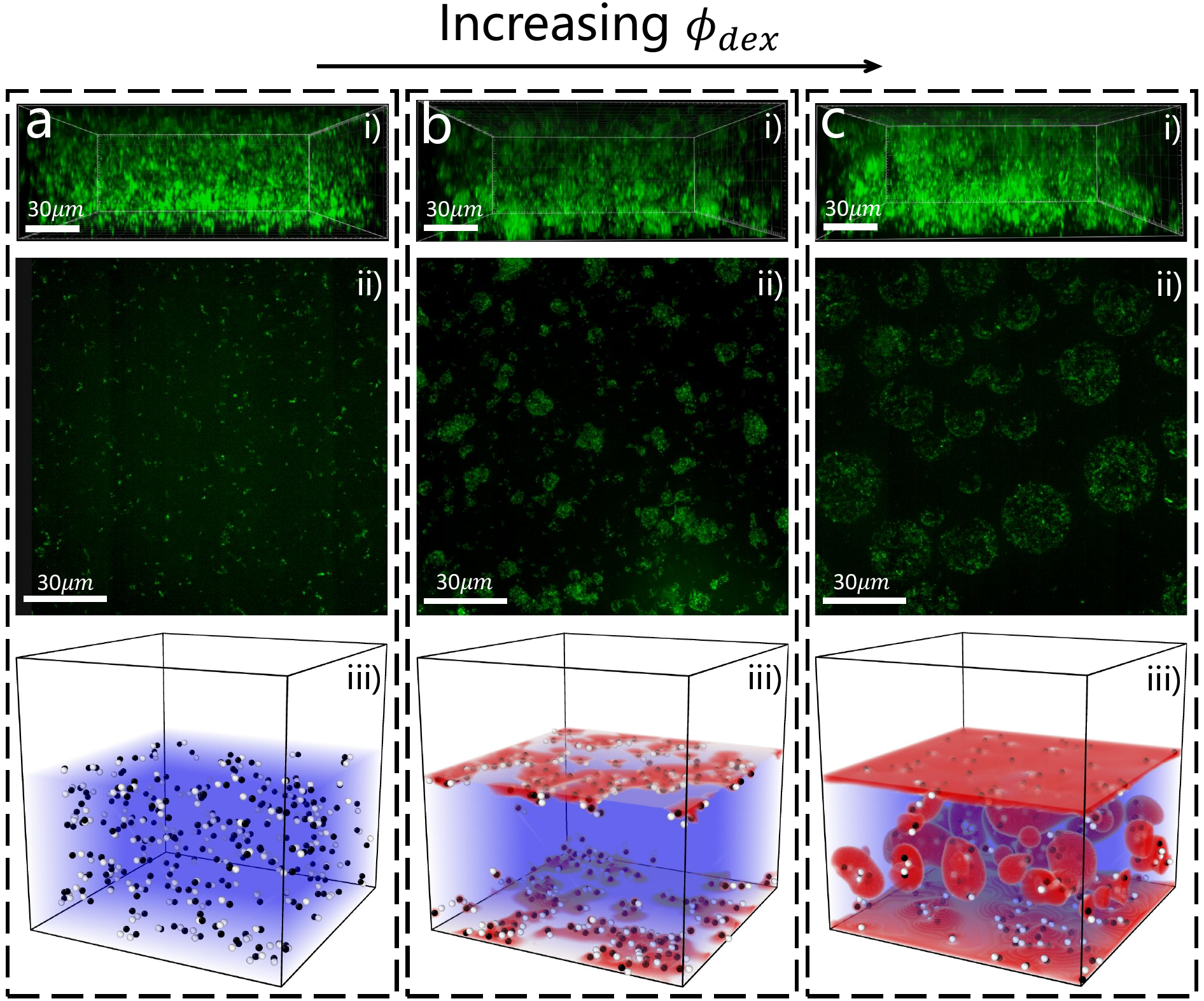}
\caption{Phase-dependent bacterial accumulation.
Confocal microscopy images of green fluorescent protein (GFP)-tagged \textit{P. aeruginosa} in binary fluids with increasing volume fraction of the DEX-rich phase. Panels (a)--(c) correspond to $\phi_{\tn{dex}}=0$, $0.05$, and $0.2$, respectively, at a fixed bacterial volume fraction of $\phi_{\tn{bac}}=0.1$. 
(i) Vertical cross-sectional views ($xz$ plane) showing the three-dimensional distribution of bacteria above the substrate. 
(ii) Horizontal views ($xy$ plane) taken within a 2.5~$\mu$m-thick layer near the substrate. 
(iii) Corresponding simulation snapshots illustrating preferential wetting of bacteria (particles) by the DEX-rich phase (red) relative to the PEG-rich phase (blue). 
(a) In the absence of DEX, bacteria remain broadly dispersed in the bulk with minimal interfacial accumulation. 
(b) At $\phi_{\tn{dex}}=0.05$, a surface-associated DEX-rich wetting layer forms, energetically trapping bacteria and leading to enhanced interfacial accumulation. 
(c) At $\phi_{\tn{dex}}=0.2$, large self-spinning droplets form in the bulk, reducing bacterial accumulation at the solid surface.
}
\label{fig:figure2}
\end{figure}

We next employ confocal imaging to directly visualize bacterial accumulation near the solid surface (Fig.~\ref{fig:figureS1}(b); see Appendix~\ref{sec:exp} for experimental details). Figure~\ref{fig:figure2} summarizes the morphological evolution of the system as the volume fraction of the DEX-rich phase, $\phi_{\text{dex}}$, is increased from $0$ to $0.2$ at a fixed bacterial volume fraction of $\phi_{\text{bac}}=0.1$. In the absence of DEX ($\phi_{\text{dex}}=0$; Fig.~\ref{fig:figure2}(a)), bacteria remain broadly dispersed in the bulk, exhibiting only weak, transient adhesion to the glass substrate after 5 minutes. Introducing a minority DEX-rich phase at $\phi_{\text{dex}}=0.05$ leads to a qualitatively different behavior: a surface-associated DEX-rich wetting layer forms, within which bacteria become confined near the interface, resulting in patchy clustering and a pronounced enhancement of interfacial accumulation (Fig.~\ref{fig:figure2}(b) and Movie~S1).

Upon further increasing the DEX-rich phase to $\phi_{\text{dex}}=0.2$, the system enters a distinct regime characterized by the formation of large DEX-rich droplets in the bulk that engulf bacteria (Fig.~\ref{fig:figure2}(c) and Movie~S2). In this high-$\phi_{\text{dex}}$ regime, interfacial adhesion remains stronger than in the single-phase control ($\phi_{\text{dex}}=0$; Fig.~\ref{fig:figure2}(a)) but is reduced relative to the maximal observed at low $\phi_{\text{dex}}=0.05$. This behavior reveals a non-monotonic dependence of bacterial accumulation on $\phi_{\text{dex}}$, with optimal adhesion occurring at intermediate phase volume fractions. The same trend is quantified in Figs.~\ref{fig:figureS2} and \ref{fig:figureS3}, which report bacterial adhesion after a fixed elapsed time of 5 minutes across different values of $\phi_{\text{dex}}$.

\subsection{Particle-field hybrid modeling}

To gain mechanistic insight into the observed behaviors, we develop a coarse-grained particle--field hybrid model to simulate active bacteria in a binary fluid under confinement. This framework captures the coupled roles of LLPS, hydrodynamic flows, bacterial motility, wetting thermodynamics, and wall boundary effects within a single computational scheme. Importantly, the model provides a general platform applicable to a broad class of microorganisms and synthetic active colloids in complex, phase-separating fluids, whose interactions are mediated by wetting phenomena. 

Each bacterium is modeled as a pusher-type swimmer consisting of a head, a tail, and a virtual particle connected to form a rigid rod (Fig.~\ref{fig:figureS6} in Appendix~\ref{sec:sim})~\cite{furukawa2014activity}. Activity is introduced through a force dipole of strength $f_{\text{act}}$ applied to the tail and the virtual particle. The surrounding binary fluid is described by a Ginzburg--Landau free-energy functional in terms of the composition order parameter $\psi$, where $\psi=1$ and $\psi=-1$ correspond to the DEX-rich and PEG-rich phases, respectively. The system is confined between two parallel walls separated by a distance much larger than the bacterial length.

The time evolution of the composition field $\psi$ obeys a conservative advection--diffusion equation coupled to the Navier--Stokes equation, which is solved using the fluid particle dynamics (FPD) method~\cite{tanaka2000,Furukawa2018,Harlow1965}. Wetting interactions between the binary fluid and both the bacterial body and the confining walls are incorporated through coupling terms characterized by affinity coefficients. Consistent with experimental measurements showing that \textit{P. aeruginosa} is amphiphilic with a slight preference for the DEX-rich phase, we set $\gamma_1=-4$ for the head and $\gamma_2=2$ for the tail~\cite{yang2025active}. To reflect the experimentally measured weak affinity of the walls for the DEX-rich phase, we set $\gamma_{\tn{wall}}=-2$ (see Appendix~\ref{sec:sim} for simulation details).

The simulations (Fig.~\ref{fig:figure2}, bottom row) qualitatively reproduce the experimentally observed morphological evolution as the average composition $\bar{\psi}$ increases from $-1$ ($\phi_{\tn{dex}}=0$) to $-0.6$ ($\phi_{\tn{dex}}=0.2$). In agreement with the confocal observation (top and middle rows of Fig.~\ref{fig:figure2}), the addition of the DEX-rich phase induces a transition from a dispersed bacterial state (Fig.~\ref{fig:figure2}(a)iii) to surface-associated clustering (Fig.~\ref{fig:figure2}(b)iii) and, at higher $\phi_{\tn{dex}}$, to droplet-dominated morphologies (Fig.~\ref{fig:figure2}(c)iii). These results demonstrate that bacterial organization is governed by the coupled effects of LLPS and wetting affinity.

Together, the combined experimental and simulation results provide a physical basis for bacterial adhesion mediated by surface-associated LLPS and wetting, in which liquid condensates act as effective capillary bridges that facilitate robust interfacial confinement.

\subsection{Motility enhances adhesion at low phase volume}

\begin{figure}[t!]
\centering \includegraphics[width=8.5cm]{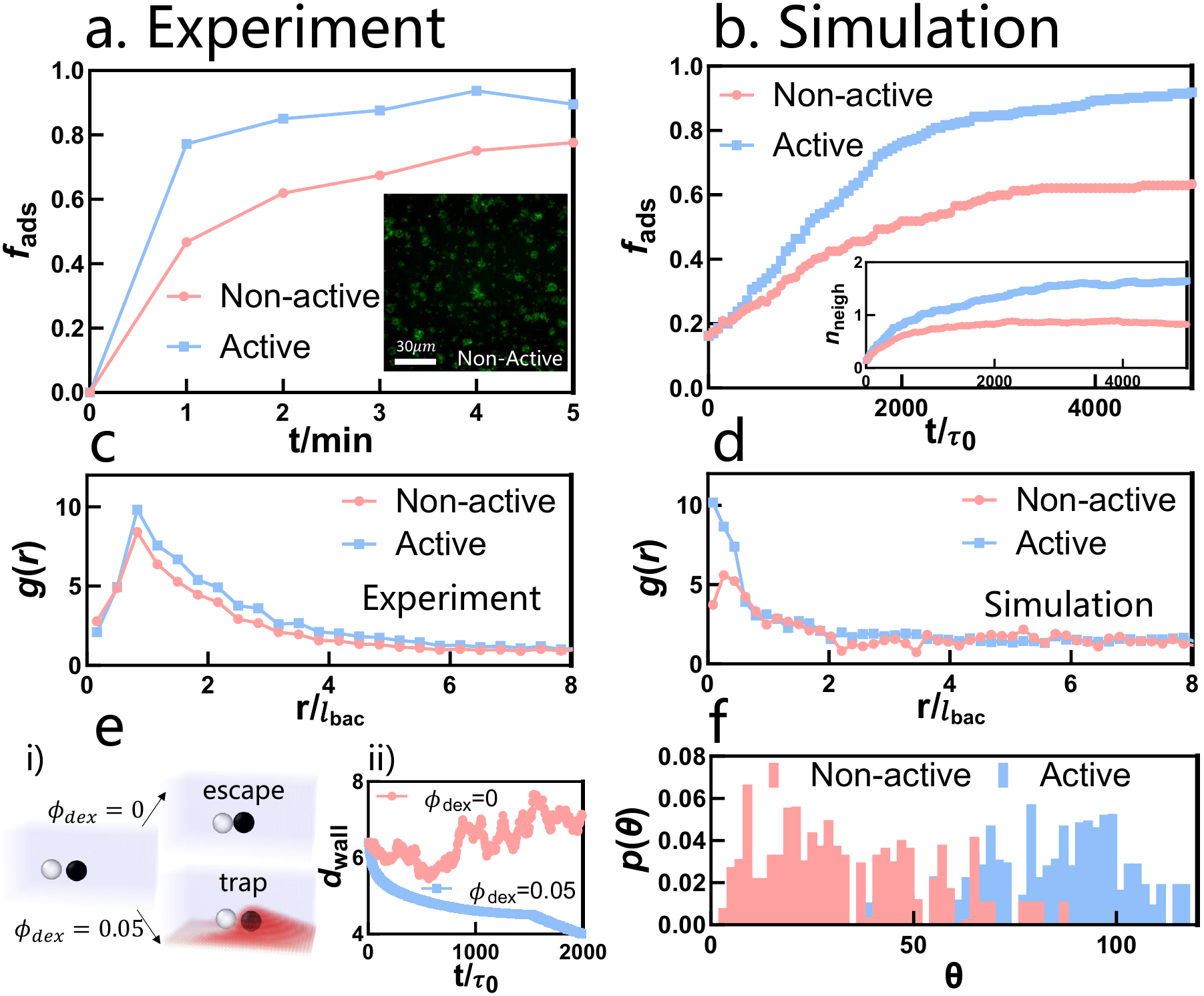}
\caption{Motility enhances interfacial accumulation at low phase volumes. 
(a,b) Temporal evolution of the accumulated bacterial fraction $f_{\tn{ads}}$ obtained from (a) experiments and (b) simulations. Active bacteria (blue) exhibit faster accumulation kinetics and reach higher steady-state levels than non-active controls (red). The inset in (a) shows a representative confocal image of the non-active bacterial layer, while the inset in (b) displays the time evolution of the average number of neighboring bacteria $n_{\tn{neigh}}$. In this study, bacteria are defined as surface-bound if they reside within a proximal layer with a thickness of 2.5~$\mu$m, corresponding to the average bacterial length.
(c,d) Radial distribution functions $g(r)$ characterizing the in-plane spatial organization of bacteria at the interface. Pronounced peaks for active bacteria in both experiments (c) and simulations (d) indicate enhanced lateral clustering relative to the non-active case. 
(e) Schematic illustration of the confinement mechanism. At $\phi_{\tn{dex}}=0.05$, the surface-associated DEX-rich wetting layer creates a deep energetic trap that confines bacteria near the interface, whereas in the single-phase control ($\phi_{\tn{dex}}=0$), bacteria readily escape into the bulk. 
(f) Probability distribution of the bacterial orientation angle $P(\theta)$. The peak near $\theta \approx 90^\circ$ indicates that confined bacteria predominantly undergo two-dimensional sliding parallel to the wall substrate.}
\label{fig:figure3}
\end{figure}

To elucidate the impact of bacterial motility on interfacial adhesion, we compared the accumulation dynamics of active and non-active \textit{P. aeruginosa} at a low DEX volume fraction of $\phi_{\text{dex}}=0.05$ (Fig.~\ref{fig:figure2}b). The time evolution of the adsorbed fraction $f_{\text{ads}}$, shown in Fig.~\ref{fig:figure3}(a) for experiments and Fig.~\ref{fig:figure3}(b) for simulations, reveals that active bacteria accumulate significantly faster and reach a higher saturation plateau than their non-motile counterparts (see also Fig.~\ref{fig:figureS4} for the mean bacterial height). This behavior demonstrates that bacterial motility enhances transport into the surface-associated DEX-rich wetting layer, thereby promoting interfacial accumulation (see Movies~S3 and S4).

For both active and non-active bacteria, the presence of a minority DEX-rich phase markedly enhances adhesion, with $f_{\text{ads}}\approx0.8$ compared to $f_{\text{ads}}\approx0.4$ in the single-phase (pure PEG solution) control after 5 minutes. Consistent with the accumulation kinetics, the average number of neighboring bacteria increases more rapidly in the active case (inset of Fig.~\ref{fig:figure3}(b)). In contrast, simulations in homogeneous fluids yield minimal accumulation ($f_{\text{ads}}<0.15$) for both active and non-active bacteria, confirming that the strong adhesion observed in simulations is not due to intrinsic bacteria--wall attraction or hydrodynamic trapping but instead arises from the LLPS-mediated mechanism (Fig.~\ref{fig:figureS5}).

To characterize the lateral organization of adsorbed bacteria, we also analyze the in-plane radial distribution function $g(r)$, shown in Fig.~\ref{fig:figure3}(c) for experiments and Fig.~\ref{fig:figure3}(d) for simulations. Active bacteria exhibit a more pronounced peak in $g(r)$ than non-active bacteria, indicating enhanced local clustering. This result shows that motility not only accelerates interfacial accumulation but also promotes in-plane aggregation, which is stabilized by capillary interactions within the wetting layer.

The physical mechanism underlying wetting-mediated adhesion is illustrated schematically in Fig.~\ref{fig:figure3}(e). In the absence of DEX ($\phi_{\text{dex}}=0$), bacteria near the wall readily escape back into the bulk under HIs (Fig.~\ref{fig:figure3}(e)i). By contrast, when a minority DEX-rich phase is present ($\phi_{\text{dex}}=0.05$), bacteria become confined within a thin DEX-rich wetting layer adjacent to the surface, as reflected by the reduced bacterium--wall distance $d_{\tn{wall}}$ (Fig.~\ref{fig:figure3}(e)ii). This wetting-mediated energetic confinement effectively suppresses bacterial escape and stabilizes adhesion.

Consistent with this physical picture, Fig.~\ref{fig:figure3}(f) shows that the wetting layer constrains active bacteria to swim predominantly parallel to the substrate, in contrast to the tilted orientations typically observed in homogeneous fluids~\cite{bianchi2017holographic}. The probability distribution $P(\theta)$ exhibits a pronounced peak near $\theta=90^\circ$, indicating planar alignment. Together, these observations demonstrate that the thin DEX-rich wetting layer acts as an effective trap for swimming bacteria, enabling robust interfacial accumulation and lateral clustering.

\subsection{Motility suppresses adhesion at high phase volume}

\begin{figure}[t!]
\centering \includegraphics[width=8.5cm]{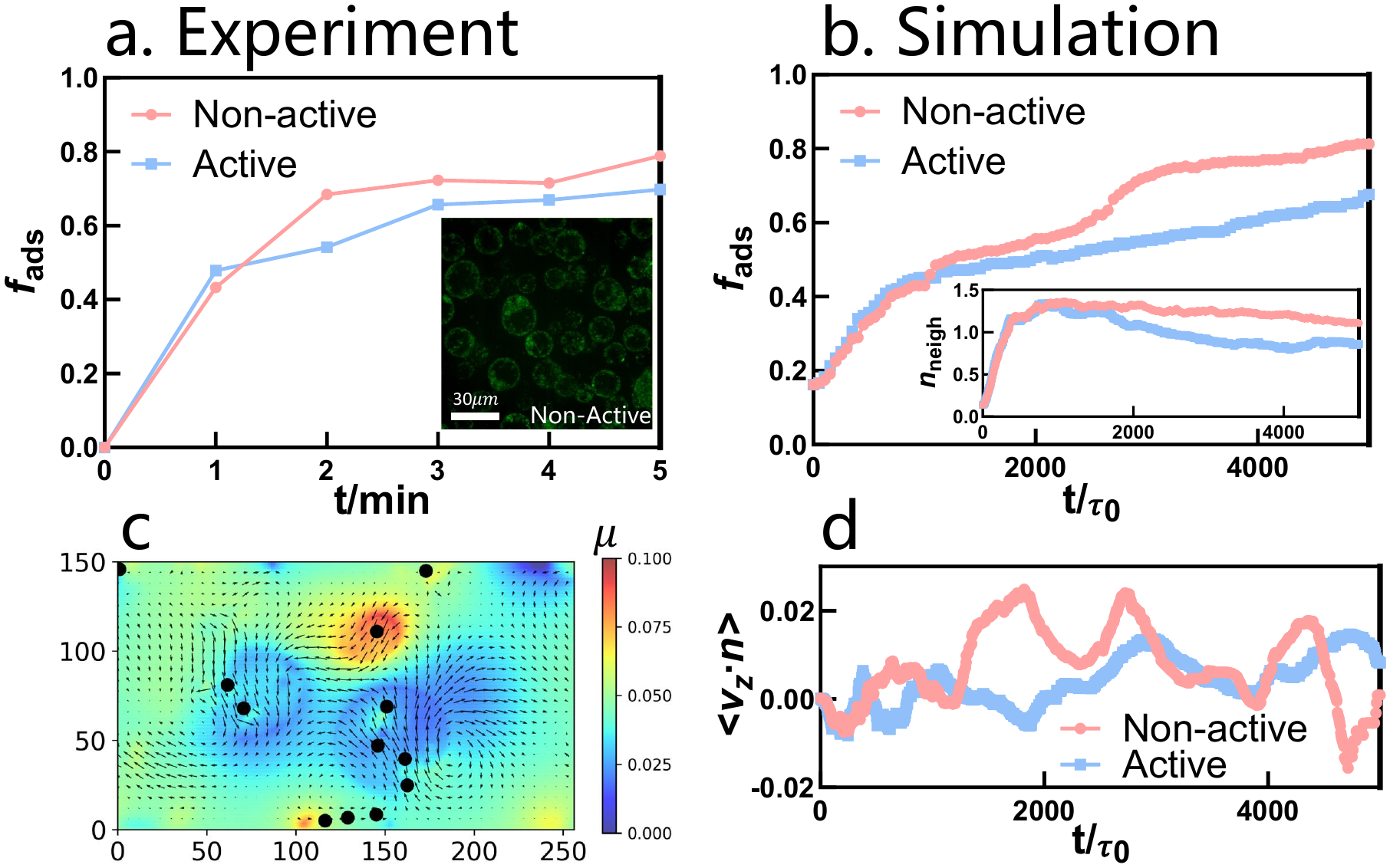}
\caption{Active motility suppresses interfacial accumulation at high phase volumes. 
(a,b) Accumulation kinetics of active (blue) and non-active (red) bacteria. In contrast to the low-$\phi_{\tn{dex}}$ regime, bacterial motility suppresses interfacial accumulation at high phase volume fractions. The inset in (a) shows dense interfacial accumulation of non-active bacteria, whereas the inset in (b) quantifies the reduced local clustering of active bacteria through the neighbor number $n_{\tn{neigh}}$. 
(c,d) Hydrodynamic suppression mechanism revealed by simulations. Panel (c) shows the flow field generated by self-spinning droplets in the bulk, which emerge from collective bacterial motion. These rotating droplets produce hydrodynamic flows that oppose the chemical-potential gradient driving transport toward the surface. As a result, the wall-normal velocity $\langle \bm v_z \cdot \bm n \rangle$ is significantly reduced for active bacteria, as shown in panel (d), leading to suppressed accumulation.}
\label{fig:figure4}
\end{figure}

\begin{figure}[t!]
\centering \includegraphics[width=8.5cm]{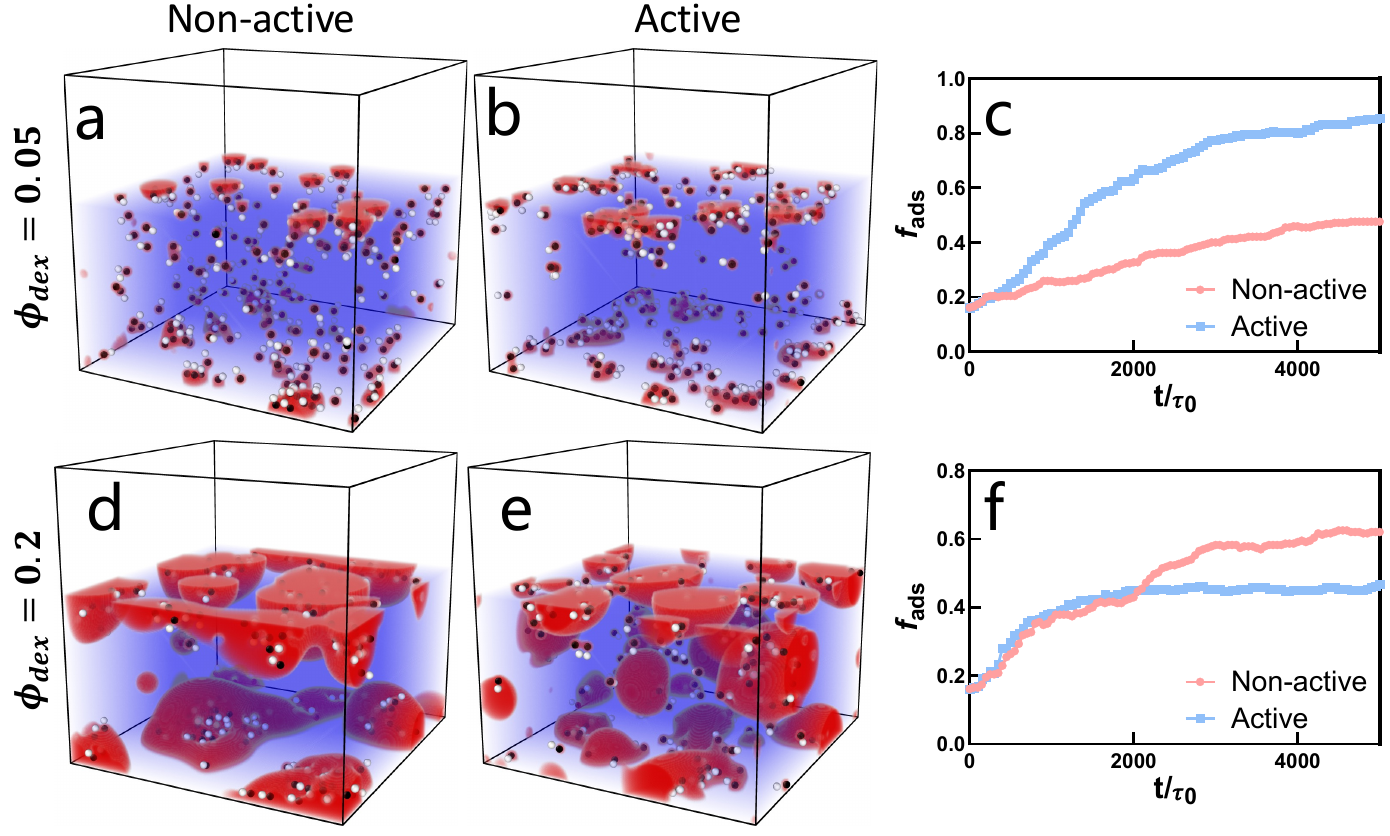}
\caption{Simulation of phase-dependent bacterial accumulation near neutral walls ($\gamma_{\tn{wall}}=0$). 
Representative simulation snapshots for low ($\phi_{\tn{dex}}=0.05$; (a,b)) and high ($\phi_{\tn{dex}}=0.2$; (d,e)) volume fractions of the minority phase. 
(c) Accumulation kinetics at $\phi_{\tn{dex}}=0.05$, demonstrating that bacterial motility promotes interfacial accumulation by accelerating transport of the wetting phase toward the walls. 
(f) Accumulation kinetics at $\phi_{\tn{dex}}=0.2$, showing that motility suppresses accumulation in this high-phase-volume regime. These results are qualitatively similar to those obtained with wetting walls.}
\label{fig:figure5}
\end{figure}

While bacterial motility might be expected to consistently enhance interfacial adhesion, we observe a qualitatively different behavior at higher DEX volume fractions ($\phi_{\text{dex}}=0.2$; Fig.~\ref{fig:figure2}(c)). In this regime, we find bacterial activity instead suppresses interfacial adsorption. This is evidenced by a lower adsorbed fraction $f_{\text{ads}}$ for active bacteria compared to non-active controls (Fig.~\ref{fig:figure4}(a) for experiments and Fig.~\ref{fig:figure4}(b) for simulations; see also Movies~S5 and S6).

We attribute this suppression to the formation of self-spinning droplets in the bulk suspension, which emerge from collective bacterial motion~\cite{dewangan2019rotating,yang2025active} (Fig.~\ref{fig:figure4}(c); see Movie~S7). These rotating droplets generate hydrodynamic flows that oppose the chemical-potential gradient ($\nabla\mu$) driving bacterial transport toward the substrate. Consistent with this picture, Fig.~\ref{fig:figure4}(d) shows a pronounced reduction in the wall-normal velocity $\langle \bm v_z \cdot \bm n \rangle$ for active bacteria. As a result, the self-spinning droplets produce an effective ``hydrodynamic lift" that inhibits interfacial accumulation.

Overall, these results demonstrate that bacterial motility plays a dual role in interfacial adhesion. At low $\phi_{\text{dex}}$, activity promotes adhesion by accelerating transport into the wetting-induced energetic trap, whereas at high $\phi_{\text{dex}}$ it suppresses adhesion through hydrodynamic lift generated by self-spinning active droplets. Consequently, adhesion efficiency depends non-monotonically on phase volume: an excess of the minority phase leads to bacterial sequestration within bulk droplets rather than enhanced surface accumulation. Although our experiments focus on wetting walls, this mechanism is generic and persists even near the ideal neutral walls ($\gamma_{\tn{wall}}=0$; Fig.~\ref{fig:figure5}).

Previous experimental studies~\cite{wu2018entrapment,wei2025confinement} have shown that bacterial accumulation in homogeneous fluids is limited, particularly for pusher-type swimmers~\cite{sartori2018wall} that readily escape hydrodynamic trapping. The wetting-mediated mechanism identified here provides a route to overcome these limitations by introducing an energetic component absent in purely hydrodynamic scenarios. This attraction is closely analogous to wetting-induced capillary interactions between passive colloids~\cite{araki2006wetting,araki2008dynamic}. More broadly, our findings offer new physical insight into bacterial aggregation~\cite{armbruster2018new,black2025capillary} and suggest a strategy for regulating surface colonization through the controlled introduction of a preferentially wetting phase.

\subsection{Generality across bacterial species}

\begin{figure}[t!]
\centering \includegraphics[width=8.5cm]{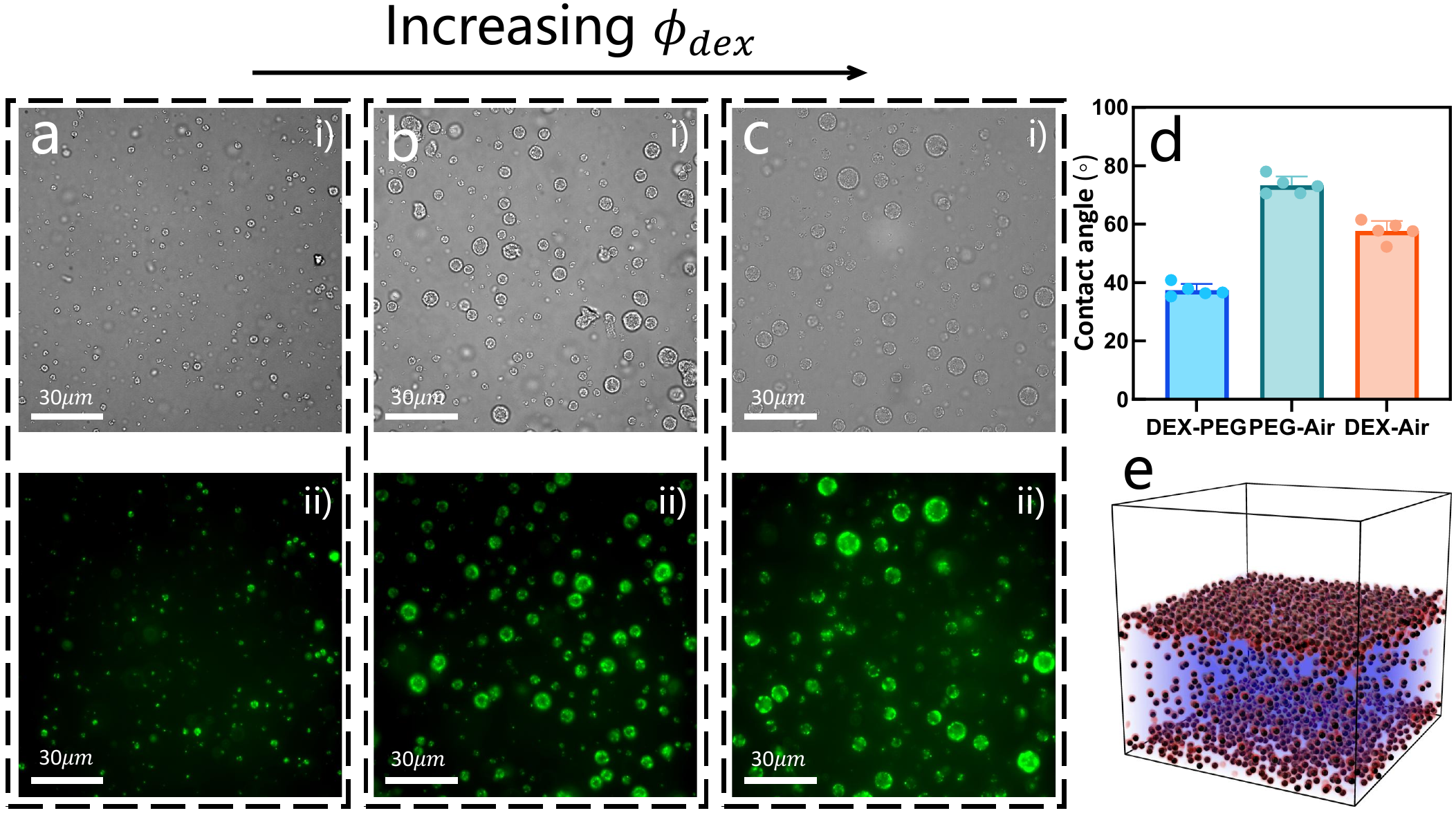}
\caption{Robust accumulation across bacterial species driven by wetting-coupled LLPS.
(a--c) Experimental images of non-motile \textit{S. aureus} in binary fluids with increasing DEX content, demonstrating that interfacial accumulation is enhanced upon the introduction of the DEX-rich phase. 
(d) Contact angle measurements of the polymer phases on a \textit{S. aureus} film, confirming a strong wetting preference for the DEX-rich phase relative to the PEG-rich phase. 
(e) Simulations showing that wetting-mediated accumulation persists for \textit{S. aureus} modeled as spherical particles.}
\label{fig:figure6}
\end{figure}

To validate the generality of the proposed mechanism, we extend our study to the non-motile bacterium \textit{S. aureus}, which exhibits an even stronger affinity for the DEX-rich phase than \textit{P. aeruginosa} (Fig.~\ref{fig:figure6}(d)). We confirm that LLPS robustly enhances the interfacial accumulation of \textit{S. aureus}, as confirmed by experiments (Fig.~\ref{fig:figure6}(a)--(c)) and FPD simulations (Fig.~\ref{fig:figure6}(e)). These results demonstrate that LLPS-mediated accumulation does not rely on bacterial motility or species-specific activity, but instead arises from generic wetting thermodynamics. Therefore, the mechanism uncovered in this study broadly applies to diverse bacterial colonization in the vicinity of interfaces.

Previous studies have reported that external shear flow can, counterintuitively, enhance interfacial adhesion of a wide range of bacterial species~\cite{thomas2002bacterial,yakovenko2008fimh,herman2018staphylococcus}. This effect is often attributed to biological ``catch-bond'' mechanisms, in which receptor-ligand bonds strengthen under tensile stress. From a physical perspective, such stress-induced strengthening can be viewed as an effective increase in attractive interactions. We hypothesize that external shear induces a local transition to a phase-separated state, analogous to shear-induced phase separation observed in polymer solutions~\cite{onuki1990shear,le2001shear,antonov2006phase}. The resulting formation of a wetting phase facilitates bacterial accumulation at interfaces, suggesting a physical ``capillary-catch" mechanism that parallels the biological ``catch-bond" explanation for shear-enhanced colonization~\cite{thomas2002bacterial,yakovenko2008fimh,herman2018staphylococcus}.

\section{Conclusion and Outlook}

In summary, by combining experiments on \textit{Pseudomonas aeruginosa} and \textit{Staphylococcus aureus} with fluid particle dynamics simulations~\cite{tanaka2000,Furukawa2018}, we have identified a general physical mechanism by which LLPS and wetting transform active matter from a dynamically biased state into an energetically confined interfacial state. Unlike hydrodynamic trapping~\cite{berke2008hydrodynamic,wu2018entrapment,wei2025confinement} or motility-induced phase separation~\cite{cates2015motility,fily2012athermal}, which rely on kinematic or collective dynamical effects without generating stable free-energy minima, wetting-coupled LLPS creates a true energetic trap at solid--liquid interfaces. This trap arises from selective partitioning into a surface-wetting phase and is accompanied by capillary-mediated attractions that promote lateral aggregation. As a result, robust surface accumulation and clustering naturally emerge even from dilute bacterial suspensions.

We demonstrate how this mechanism operates in non-equilibrium active systems. When the phase preferentially wetted by bacteria forms a surface-associated wetting layer, interfacial free-energy minimization stabilizes bacterial adhesion. We further uncover a dual role of bacterial motility in this process. At low phase volumes, activity enhances transport into the wetting layer and accelerates accumulation, whereas at higher phase volumes it suppresses adhesion through the formation of self-spinning active droplets in the bulk region that generate long-range hydrodynamic flows opposing wetting-driven transport. The simulations reproduce these morphological transitions without invoking explicit adhesive interactions, confirming that bacterial adhesion kinetics are governed by the interplay between motility, LLPS, and wetting thermodynamics. The persistence of enhanced accumulation for non-motile \textit{S. aureus} and near-neutral walls underscores the generic nature of this mechanism across bacterial species and surface properties.

Controlling biofilm formation remains a formidable challenge~\cite{brading1995dynamics}. From a biological perspective, our results suggest that LLPS provides a powerful physical route by which secreted phase-separating biopolymers can facilitate early-stage bacterial colonization and biofilm initiation~\cite{arciola2018implant,matz2005biofilm,djermoun2025biofilm}. By creating surface-associated wetting layers that generate deep energetic traps, LLPS can strongly enhance bacterial adhesion and lateral clustering already under dilute conditions. In this sense, the earliest steps of colonization---and potentially infection---may be strongly influenced by the phase behavior of polymer-rich environments. Rather than being dictated solely by biochemical specificity, bacterial adhesion emerges here as a tunable physical process controlled by interfacial energetics and active transport. Our research offers a minimal, physics-based explanation for how dense bacterial aggregates can rapidly form at interfaces despite electrostatic repulsion and weak intrinsic bacteria-wall attraction.

Beyond biology, our mechanism has broad implications for transport and surface colonization in complex fluids. In macromolecule-rich industrial and environmental settings, shear, thermal, or pH gradients may trigger LLPS near pipeline walls~\cite{indulkar2015ph,zhang2020prediction,hester2023fluid}, generating polymer-rich wetting layers that act as an effective ``capillary glue'' anchoring microorganisms to surfaces. Our findings thus provide a physical basis for the commonly observed biofouling phenomena in industrial conduits~\cite{silhan2006effect,mahapatra2015study,giorgi2022real}. More generally, wetting-coupled LLPS emerges as a unifying principle for interfacial organization in phase-separated active systems, with relevance extending to synthetic swimmers~\cite{zhang2017active}, colloidal suspensions~\cite{tanaka1994pattern,beysens1999wetting}, and smart active materials~\cite{weldrick2021smart,liese2025chemically}. We anticipate that controlling phase behavior and surface wettability will open new avenues for regulating biofouling and interfacial organization in both living and engineered active matter systems.

\begin{acknowledgments}
J.Y. acknowledges support from the startup fund of HKUST(GZ) and the National Natural Science Foundation of China (Grant No. 22503076). H.T. is supported by the Grant-in-Aid for Specially Promoted Research (JSPS KAKENHI Grant No. JP20H05619). C.W. acknowledges support from the Science and Technology Development Fund, Macao SAR (FDCT, No. 0001/2021/AKP, 0024/2023/AFJ, 0209/2024/AGJ, 0031/2023/ITP1, and 005/2023/SKL, 0002/2025/NRP); the National Natural Science Foundation of China (Grant Nos. 32361163656, 32022088, 32230056); the University of Macau (MYRG-GRG2023-00136-ICMS-UMDF, MYRG-GRG2024-00189-ICMS-UMDF, and MYRG-CRG2023-00009-IAPME); and the Zhuhai UM Science \& Technology Research Institute (CP-102-2024). Simulations were performed on the HPC platform at HKUST(GZ).
\end{acknowledgments}

\vspace{1cm}
\appendix
\centerline{\bf Appendix}
\renewcommand{\thefigure}{A\arabic{figure}}
\setcounter{figure}{0}   

\section{Experimental Method}
\label{sec:exp}

\subsection{Bacterial culture and staining}

The bacterial strains used in this study were \textit{Pseudomonas aeruginosa} (\textit{P. aeruginosa}; ATCC 10145), its green fluorescent protein (GFP)-tagged variant (ATCC 10145GFP), and \textit{Staphylococcus aureus} (\textit{S. aureus}; ATCC 25923). Bacterial stocks were stored at $-80^\circ$C in Luria--Bertani (LB) broth (Solarbio, L8290). For routine experiments, bacteria were streaked from frozen stocks onto LB agar plates (Solarbio, L8291) and incubated at $37^\circ$C for 24~h.

For liquid cultures, a single colony was inoculated into 10~mL of LB broth (Solarbio, L8290) and grown overnight at $37^\circ$C with shaking at 150~rpm. This overnight culture was then diluted into fresh LB medium and grown to mid-exponential phase, corresponding to an optical density at 600~nm of $\mathrm{OD}_{600}=0.5$--0.6. All culture media and reagents were sterilized by autoclaving prior to use.

\paragraph{Fluorescent staining of \textit{S. aureus}.}
Unlike GFP-tagged \textit{P. aeruginosa}, \textit{S. aureus} cells were fluorescently labeled using the CellTrace$^{\mathrm{TM}}$ CFSE Cell Proliferation Kit (Thermo Fisher Scientific, C34554). Harvested bacteria were washed with phosphate-buffered saline (PBS) and resuspended in PBS containing 5~$\mu$M CFSE. The suspension was incubated at $37^\circ$C for 20~min in the dark. Following incubation, the cells were washed three times with fresh LB medium to remove excess dye before use in subsequent experiments.

\paragraph{Biomass quantification.}
Bacterial biomass was quantified using the packed cell volume (PCV) method. Aliquots of 500~$\mu$L of bacterial culture, collected at specified time points, were transferred to calibrated 1.5~mL conical centrifuge tubes (Axygen, MCT-150-C-S) and centrifuged at $5{,}000\times g$ for 15~min at room temperature. After centrifugation, the supernatant was carefully removed, and the volume of the resulting cell pellet was read directly from the tube graduations. The PCV was calculated as the ratio of pellet volume to the initial culture volume, expressed as a percentage ($V_{\text{pellet}}/V_{\text{total}}\times100$). All measurements were performed in triplicate.

\subsection{Preparation and inoculation of aqueous two-phase systems (ATPS)}

\paragraph{Media preparation.}
Stock solutions of dextran (DEX; molecular weight 450--650~kDa; Sigma-Aldrich, 31392-50G) and polyethylene glycol (PEG; molecular weight 100~kDa; Sigma-Aldrich, U07G051-100G) were prepared by dissolving each polymer in standard Luria--Bertani (LB) broth (Solarbio, L8290) to a final concentration of 5~wt\%. The solutions were sterilized by autoclaving and stored at $4^\circ$C prior to use.

\paragraph{Bacterial strain adaptation.}
Prior to inoculation into the ATPS, \textit{P. aeruginosa} (ATCC 10145) and \textit{S. aureus} (ATCC 25923) were adapted to the polymer-rich environment. A single colony was first grown overnight in standard LB at $37^\circ$C with shaking at 150~rpm. The culture was then pelleted by centrifugation (5{,}000~$\times g$, 10~min), the supernatant was discarded, and the cells were resuspended in sterile DEX--LB solution (5~wt\%). The suspension was incubated overnight at $37^\circ$C with shaking (150~rpm) to allow metabolic adaptation to the polymer-rich medium.

\paragraph{Biomass quantification and inoculation.}
Following adaptation, the bacterial cultures were centrifuged (5{,}000~$\times g$, 15~min, $4^\circ$C), and the supernatant was removed. The total bacterial biomass was quantified using the packed cell volume (PCV) method described above.

\paragraph{ATPS assembly.}
ATPS samples were assembled in sterile 1~mL PCR tubes (Vazyme, PCR00802). The pre-adapted bacterial pellet (\textit{P. aeruginosa} or \textit{S. aureus}) was first resuspended in the 5~wt\% DEX--LB stock solution, followed by the addition of the 5~wt\% PEG--LB stock solution. The components were mixed by gentle pipetting using 200~$\mu$L tips to obtain the desired final volume fractions of bacteria ($\phi_{\text{bac}}$), the DEX-rich phase ($\phi_{\text{dex}}$), and the PEG-rich phase ($\phi_{\text{peg}}$). The total sample volume was fixed at 100~$\mu$L.

\subsection{Bright-field observation, confocal microscopy, and three-dimensional imaging}

After phase separation, ATPS samples containing GFP-tagged \textit{P. aeruginosa} (ATCC 10145GFP) or CFSE-labeled \textit{S. aureus} were gently transferred to 35~mm glass-bottom dishes (Nest, 801001). To minimize evaporation and preserve the biphasic structure, imaging was performed in a humidified chamber maintained at a relative humidity exceeding 90\%.

Imaging was carried out using an Olympus Spin SR microscope equipped with a Yokogawa CSU-W1 spinning-disk confocal unit and a $100\times$/1.4~NA oil-immersion objective. To resolve the spatiotemporal organization of bacteria, four-dimensional ($xyzt$) time-lapse sequences were acquired. The FITC channel was used to visualize both GFP-expressing \textit{P. aeruginosa} and CFSE-labeled \textit{S. aureus}. Three-dimensional reconstructions were obtained from $z$-stacks acquired with a step size of 1~$\mu$m.

\subsection{Image analysis and quantification of bacterial adhesion}

To quantify the spatiotemporal dynamics of bacterial adhesion from four-dimensional ($x,y,z,t$) confocal datasets, we develop an automated analysis pipeline in Python using the \texttt{scikit-image} and \texttt{scipy} libraries.

\paragraph{Preprocessing and three-dimensional localization.}
Raw image stacks were first rescaled to reduce computational cost. To enhance bacterial contrast and correct for uneven illumination, a white top-hat transform (disk structuring element) was applied to each $z$ slice, followed by Gaussian smoothing ($\sigma=1.0$) for noise suppression. Bacterial centroids were then identified in three dimensions using a Laplacian-of-Gaussian (LoG) blob detection algorithm with an adaptive detection threshold defined by local background statistics, $\mu_{\mathrm{bg}}+3\sigma_{\mathrm{bg}}$.

\paragraph{Dynamic surface localization and region definition.}
The axial position of the substrate~$Z_{\mathrm{surf}}$ was determined independently for each time frame by constructing the $z$-position histogram of all detected bacterial centroids. The peak of this distribution was taken as the surface plane, and a temporal exponential smoothing filter ($\alpha=0.2$) was applied to suppress frame-to-frame jitter. Based on $Z_{\mathrm{surf}}$, two spatial regions were defined: (i) an \emph{adhesion layer}, defined as a 5.0~$\mu$m-thick slab centered on the surface ($Z_{\mathrm{surf}}\pm2.5~\mu$m); and (ii) an \emph{effective pool}, defined as a 30.0~$\mu$m-thick slab centered on the surface ($Z_{\mathrm{surf}}\pm15.0~\mu$m), representing the bulk bacterial population available for adsorption. Note that while bacteria are restricted to the upper half-space, the lower half-space is included to account for axial fluorescence smearing, thus preventing bacterial undercounting during 3D detection.

\paragraph{Background exclusion and kinetic quantification.}
To distinguish newly adsorbed bacteria from pre-existing surface-associated cells, a spatial exclusion procedure was implemented. Bacteria located within the adhesion layer at $t=0$ were labeled as background. In subsequent frames, any particle detected within a lateral radius of 3.0~$\mu$m from a background coordinate was excluded from the analysis. The adsorbed fraction was then defined as
\begin{equation}
f_{\tn{ads}}=\frac{N_{\tn{ads}}}{N_{\tn{pool}}},
\end{equation}
where $N_{\tn{ads}}$ is the number of newly adhered bacteria within the adhesion layer and $N_{\tn{pool}}$ is the number of bacteria within the effective pool.

\paragraph{Spatial organization.}
The in-plane spatial organization of adhered bacteria was characterized by computing the radial distribution function (RDF) $g(r)$. For particles near the boundaries of the field of view, standard RDF estimators suffer from finite-size bias because the sampling annulus partially extends beyond the imaged area. To correct for this effect, the measured RDF was normalized by that of an ideal gas, obtained from Monte Carlo simulations of randomly distributed particles within the same imaging geometry.

\section{Simulation method}
\label{sec:sim}

\subsection{Hybrid model for bacteria in a binary fluid}

We propose a particle--field hybrid framework to investigate the non-equilibrium accumulation dynamics of bacteria suspended in a phase-separating binary fluid near confining walls. The model simultaneously captures phase separation, hydrodynamic interactions, active self-propulsion, wetting thermodynamics, and wall confinement within a unified description. Importantly, this framework is readily extensible to other microorganisms and synthetic active colloids whose interactions with multicomponent fluids are governed by wettability.

\paragraph{Representation of a bacterium.}
To model pusher-type \textit{P. aeruginosa}, we follow Ref.~\cite{furukawa2014activity} and represent each bacterium as a trimer of particles (Fig.~\ref{fig:figureS6}). Specifically, bacterium $i$ consists of a head particle of radius $a_{\tn{h}}$ located at $\bm r_i^{\mathrm{h}}$, a tail particle of radius $a_{\tn{t}}$ at $\bm r_i^{\mathrm{t}}$, and a virtual particle of radius $a_{\tn{vir}}$ at $\bm r_i^{\mathrm{vir}}$. The head and tail particles are connected by a harmonic spring of equilibrium length $\ell_1$, which maintains the elongated geometry of the bacterial cell body.

The position of the virtual particle is defined relative to the tail--head orientation vector
$$\hat{\bm D} = \frac{\bm r_i^{\mathrm{h}} - \bm r_i^{\mathrm{t}}}{\lvert \bm r_i^{\mathrm{h}} - \bm r_i^{\mathrm{t}} \rvert},$$
such that $\bm r_i^{\mathrm{vir}} = \bm r_i^{\mathrm{t}} - \ell_2 \hat{\bm D}$. Bacterial activity is introduced by applying a constant propulsion force of magnitude $f_{\tn{act}}$ to the tail particle in the direction of the head, while an equal and opposite force is applied to the virtual particle. This force dipole generates the characteristic dipolar flow field of a pusher-type swimmer~\cite{drescher2011fluid} while conserving total momentum.

To model the non-motile, approximately spherical bacterium \textit{S. aureus}, each cell is represented as a single spherical particle of radius $a_{\tn{h}}$.

\paragraph{Fluid particle dynamics.} 
To capture the long-range HIs induced by bacterial activity, we solve the Navier--Stokes (NS) equation using the fluid particle dynamics (FPD)  method~\cite{tanaka2000,Furukawa2018,yuan2025network}. In the FPD framework, each constituent particle of a bacterium is treated as a viscous inclusion with a diffuse interface, described by the scalar field
\begin{equation}
\phi_{i}({\bm r})=\frac{1}{2}\left\{\tanh
\left[\frac{a_i-\left|{\bm r}-{\bm r}_{i}\right|}{\xi}\right] + 1\right\},
\label{eq:phi}
\end{equation}
where $a_i$ is the radius of particle $i$, $\bm r_i$ is its center position, and $\xi$ denotes the thickness of the diffuse interface.

The spatially varying viscosity field is then defined as
\begin{equation}
\eta(\bm{r}) = \eta_{\rm s} + \left(\eta_{\rm p}-\eta_{\rm s}\right)\sum_{i=1}^{N_\tn p} \phi_i(\bm{r}),
\label{eq:eta}
\end{equation}
where $\eta_{\rm s}$ and $\eta_{\rm p}$ are the viscosities of the solvent and particle phases, respectively, and $N_\tn p$ is the total number of constituent particles.  Specifically, $N_{\tn p}=2N_{\tn b}$ for \textit{P. aeruginosa} (head and tail particles) and $N_{\tn p}=N_{\tn b}$ for \textit{S. aureus}, where $N_{\tn b}$ denotes the total number of bacteria. 

By choosing a large viscosity contrast, $\eta_{\rm p}=50\,\eta_{\rm s}$, fluid motion inside the particles is strongly suppressed, such that each particle behaves effectively as a rigid body while remaining hydrodynamically coupled to the surrounding fluid~\cite{tanaka2000,Furukawa2018}.

\paragraph*{Total free energy.}
The binary fluid with embedded bacteria is governed by the total free energy functional~\cite{araki2006wetting,Furukawa2018}
\begin{equation}
\begin{aligned}
\mathscr{F}\{\psi, \bm{r}_i\}
&= \mathscr{F}_b\{\psi\} + \mathscr{F}_s\{\psi, \bm{r}_i\} + \mathscr{F}_a\{\psi, \bm{r}_i\} \\
&\quad + \mathscr{F}_{\rm wall} + \mathscr{F}_{\rm wet} + U\{\bm{r}_i\} + U_\tn{wall}\{\bm{r}_i\}
\end{aligned}
\label{eq:F}
\end{equation}
where $\psi(\bm{r})$ denotes the composition order parameter of the binary mixture and $\bm{r}_i$ represents the positions of the bacterial constituent particles. Each term accounts for a distinct physical contribution, including bulk phase separation, particle--fluid coupling, activity-related interactions, wetting effects, and steric interactions with other particles and the confining walls.

\paragraph*{Bulk free energy.}
The bulk contribution of the binary fluid mixture is described by a standard Ginzburg--Landau free-energy functional,
\begin{equation}
\mathscr{F}_b\{\psi\} = \int d\bm{r} \left[  \frac{1}{2}r_0 \epsilon_0 \psi^2 + \frac{1}{4}u_0 \psi^4 + \frac{k_0}{2}|\nabla \psi|^2 \right],
\label{eq:Fb}
\end{equation}
where $\epsilon_0 = (T-T_{\rm c})/T_{\rm c}$ is the reduced temperature relative to the critical temperature $T_{\rm c}$.
For $r_0\epsilon_0 < 0$ and $u_0>0$, the homogeneous state becomes unstable and the mixture undergoes phase separation into two coexisting phases, $A$ and $B$, with equilibrium order parameters
\begin{equation}
\psi_{\rm equi}^{A,B} = \pm\sqrt{-\frac{r_0\epsilon_0}{u_0}}.
\end{equation}

\paragraph*{Surface free energy.}
Wetting interactions between the particles and the binary fluid are incorporated through a surface free-energy contribution, $\mathscr{F}_{\rm s}\{\psi,\bm r_i\}$,
\begin{equation}
\mathscr{F}_s\{\psi,\mathbf{r}_i\} = \sum_{i=1}^{N_\tn p} \int d\mathbf{r}\,|\nabla\phi_i|\, f_s(\psi,T),
\quad f_s(\psi,T)=\gamma_i\psi.
\end{equation}
Here $\gamma_i$ is the wetting affinity parameter associated with that particle. 

The sign of $\gamma_i$ determines the preferred phase: $\gamma_i>0$ favors wetting by phase~B, whereas $\gamma_i<0$ favors wetting by phase~A. To capture the amphiphilic nature of \textit{P. aeruginosa}, different affinities are assigned to the head and tail particles, with $\gamma_1=-4$ for the head and $\gamma_2=2$ for the tail. This asymmetric wettability promotes the accumulation of the bacterium at fluid interfaces. In contrast, non-motile \textit{S. aureus} is modeled as a single spherical particle, for which we set a uniform wetting parameter $\gamma_{\tn{sa}}=-4$, corresponding to a preference for the DEX-rich phase.

\paragraph*{Particle impermeability.}
To enforce the near-impermeability of bacterial bodies to the binary fluid components, we introduce an additional free-energy contribution,
\begin{equation}
\mathscr{F}_a\{\psi,\bm{r}_i\} = \sum_{i=1}^{N_\tn p} \int d\bm{r}\,\chi \phi_i \psi^2,
\label{eq:Fa}
\end{equation}
with $\chi>0$ penalizing spatial variations of the composition field within the particle volumes. This term effectively suppresses phase separation inside the bacterial bodies, enforcing $\psi \approx 0$ within the head and tail regions and ensuring that compositional variations are confined to the surrounding fluid.

\paragraph*{Wall confinement.}
The bacterial suspension and the binary fluid mixture are confined between two parallel walls located at $z=0$ and $z=H$. The walls are represented by a continuous field, $\phi_{\rm wall}(z)=\tfrac{1}{2}\bigl[\tanh((L-H)/2-|z-(L+H)/2|)/\xi)+1\bigr]$,  
which defines a high-viscosity region that effectively acts as a rigid boundary.

Analogous to the particle-impermeability constraint, we suppress compositional variations inside the wall region by introducing the free-energy term
\begin{equation}
\mathscr F_{\rm wall}=\int d\bm r\,\chi_{\rm wall}\,\phi_{\rm wall}\psi^2,
\end{equation}
which enforces $\psi \approx 0$ within the walls.  

The preferential wetting of the binary fluid components at the walls is captured by an additional surface free-energy contribution,
\begin{equation}
\mathscr F_{\rm wet}=\int d\bm r\,|\nabla\phi_{\rm wall}|\,\gamma_{\rm wall}\psi ,
\end{equation}
where $\gamma_{\rm wall}$ controls the wetting affinity of the walls. Positive and negative values of $\gamma_{\rm wall}$ correspond to preferential wetting by the two phases, while $\gamma_{\rm wall}=0$ represents a neutral wall.

\paragraph*{Direct particle interactions.}
Rather than introducing an ad hoc rule that suppresses propulsion when bacteria approach one another, we explicitly include steric interactions involving the virtual site to prevent unphysical overlaps~\cite{furukawa2014activity}. In this formulation, the virtual particle is no longer purely ``phantom'' but participates in excluded-volume interactions. This treatment allows the active propulsion force to remain operative even in dense configurations, while the strong repulsion acting on the virtual site prevents overlaps between bacterial bodies and flagella.

The total direct interaction potential consists of excluded-volume repulsion between all constituent particles (head, tail, and virtual sites), together with a harmonic bond that maintains the head--tail separation,
\begin{equation}
U\{\bm{r}_i\}=\sum_{i=1}^{3 N_\tn b}\sum_{j\ne i}U_{\rm rep}(r_{ij}) + \sum_\alpha^{N_\tn b} U_{\rm bond}\left(\left|\boldsymbol{r}_\alpha^{\mathrm{h}}-\boldsymbol{r}_\alpha^{\mathrm{t}}\right|\right),
\end{equation}
where $N_{\tn b}$ is the total number of bacteria, and each bacterium contributes three particles (head, tail, and virtual).

The excluded-volume interaction is modeled by a steep repulsive potential,
\begin{equation}
U_{\rm rep}(r_{ij})=\epsilon_1\left(\frac{2a_{ij}}{r_{ij}}\right)^{24},
\label{eq:WCA}
\end{equation}
and the head--tail bond is described by a harmonic potential,
\begin{equation}
U_{\rm bond}(r_{ij})=\frac{1}{2} \epsilon_2\left(\frac{r_{ij}}{\ell_1}-1\right)^2.
\end{equation}
Here $r_{ij}=\lvert \bm r_i-\bm r_j\rvert$ is the separation between particles $i$ and $j$, $a_{ij}=(a_i+a_j)/2$ is the mean particle radius, $\epsilon_1$ sets the strength of the steric repulsion, and $\epsilon_2$ is the spring constant of the head--tail bond of equilibrium length $\ell_1$.
The upper index $3N_\tn{p}$ reflects the fact that each bacterium contributes a head, a tail, and a virtual particle. 

For non-motile \textit{S. aureus}, modeled as single spherical particles, direct interactions reduce to excluded-volume repulsion only,
\begin{equation}
U\{\bm r_i\}
=\sum_{i=1}^{N_{\tn b}}\sum_{j\neq i} U_{\rm rep}(r_{ij}),
\end{equation}
which prevents overlap between bacterial bodies.

\paragraph*{Particle--wall interactions.}

Steric repulsion between particles and the confining walls is included to prevent unphysical overlap. Particle--wall interactions are modeled using the Weeks--Chandler--Andersen (WCA) potential,
\begin{equation}
U_{\rm wall}(z_i)=
\left\{
\begin{array}{l}
4\varepsilon\!\left[\left(\frac{a_i}{z_i}\right)^{12}
-\left(\frac{a_i}{z_i}\right)^{6}
+\tfrac14\right],\\[4pt]
\qquad \tn{for } z_i \le 2^{1/6}a_i \\[8pt]
4\varepsilon\!\left[\left(\frac{a_i}{(H-z_i)}\right)^{12}
-\left(\frac{a_i}{(H-z_i)}\right)^{6}
+\tfrac14\right],\\[4pt]
\qquad \tn{for } z_i \ge H-2^{1/6}a_i \\[8pt]
0, \\[4pt]
\qquad \text{otherwise},
\end{array}
\right.
\label{eq:Uwall}
\end{equation}
where $z_i$ denotes the vertical position of particle $i$, $a_i$ is its radius, $H$ is the wall separation, and $\varepsilon$ sets the strength of the repulsive interaction. This purely repulsive potential ensures excluded-volume interactions with both walls while allowing free motion in the interior of the channel.

\paragraph*{No-slip boundary conditions.}

Following the approach in previous studies~\cite{nakayama2005simulation,tateno2024mechanical}, we impose no-slip boundary conditions by introducing a frictional force density that damps fluid motion in the vicinity of the walls,
\begin{equation}
\bm{f}_{\rm fric}(\bm{r})
= -\zeta_{\rm w}\,\phi_{\rm w}(z)\,\bm{v}(\bm{r})
\label{eq:fric_density_clean}
\end{equation}
where $\bm v(\bm r)$ is the local fluid velocity, $\phi_{\rm w}(z)$ is the smooth wall indicator field, and $\zeta_{\rm w}$ is the wall friction coefficient. This term selectively suppresses fluid motion within the wall region, thereby effectively enforcing no-slip boundary conditions at the confining surfaces while leaving the bulk hydrodynamics unaffected.

\paragraph*{Force transformation.}

For the three-bead swimmer model used to represent \textit{P. aeruginosa}, the virtual site does not constitute an independent dynamical degree of freedom. Consequently, any interaction force evaluated at the virtual-site position ($\bm r_i^{\mathrm{vir}}$),
$\bm f_i^{\mathrm{vir}} = -{\partial ({U}+U_\tn{wall}})/{\partial \bm r_i^{\mathrm{vir}}}$,
must be redistributed onto the physical degrees of freedom, namely the head and tail particles of bacterium~$i$.

Using the chain rule, the force contributions transmitted to the head and tail are given by
\begin{equation}
\bm f_i^{(\mathrm{h,vir})} = -\frac{\partial {U}}{\partial \bm r_i^{\mathrm{vir}}}
\frac{\partial \bm r_i^{\mathrm{vir}}}{\partial \bm r_i^{\mathrm{h}}}
\end{equation}
and
\begin{equation}
\bm f_i^{(\mathrm{t,vir})}
= -\frac{\partial {U}}{\partial \bm r_i^{\mathrm{vir}}}
\frac{\partial \bm r_i^{\mathrm{vir}}}{\partial \bm r_i^{\mathrm{t}}}.
\end{equation}
The derivatives of the virtual-site position with respect to the head and tail coordinates are
\begin{equation}
\frac{\partial \bm r_i^{\mathrm{vir}}}{\partial \bm r_i^{\mathrm{h}}}
= - \frac{\ell_2}{|\bm r_i^{\mathrm{h}} - \bm r_i^{\mathrm{t}}|}
(\bm I - \hat{\bm D}_i \hat{\bm D}_i),
\end{equation}
and
\begin{equation}
\frac{\partial \bm r_i^{\mathrm{vir}}}{\partial \bm r_i^{\mathrm{t}}}
= \bm I +  \frac{\ell_2}{|\bm r_i^{\mathrm{h}} - \bm r_i^{\mathrm{t}}|}
(\bm I - \hat{\bm D}_i \hat{\bm D}_i),
\end{equation}
where $\hat{\bm D}_i=(\bm r_i^{\mathrm{h}}-\bm r_i^{\mathrm{t}})/\lvert \bm r_i^{\mathrm{h}}-\bm r_i^{\mathrm{t}} \rvert$ is the orientation unit vector and $\bm I$ denotes the identity tensor.
 
Substituting these expressions yields the effective forces acting on the head and tail particles,
\begin{equation}
\bm f_i^{(\mathrm{h,vir})}
= - \frac{\ell_2}{|\bm r_i^{\mathrm{h}} - \bm r_i^{\mathrm{t}}|}
(\bm I - \hat{\bm D}_i \hat{\bm D}_i)\,\bm f_i^{\mathrm{vir}}
\end{equation}
and
\begin{equation}
\bm f_i^{(\mathrm{t,vir})}
= \bm f_i^{\mathrm{vir}} + \frac{\ell_2}{|\bm r_i^{\mathrm{h}} - \bm r_i^{\mathrm{t}}|}
(\bm I - \hat{\bm D}_i \hat{\bm D}_i)\,\bm f_i^{\mathrm{vir}}.
\end{equation}
By construction, these forces satisfy $\bm{f}_i^{(\mathrm{h,vir})} + \bm{f}_i^{(\mathrm{t,vir})} = \bm{f}_i^{\mathrm{vir}}$, ensuring that the interaction force acting at the virtual site is conservatively distributed between the head and tail without introducing any spurious net force on the bacterium.

\paragraph*{Navier-Stokes equation and hydrodynamics.}
The fluid velocity field $\bm{v}$ obeys the Navier-Stokes (NS) equation,
\begin{equation}
\rho\left(\frac{\partial}{\partial t} + \bm{v}\cdot\nabla\right)\bm{v}
= \bm{f}_{\rm rev} + \nabla\cdot\bm{\sigma},
\label{eq:NS}
\end{equation}
where $\rho$ is the fluid mass density. The viscous stress tensor is given by
\begin{equation}
\bm{\sigma} = \eta(\bm{r})\left[\nabla\bm{v}+(\nabla\bm{v})^\mathsf{T}\right] - p\bm{I}
\end{equation}
with $\eta(\bm r)$ the position-dependent viscosity, $\bm I$ the identity tensor, and the pressure $p$ determined so as to enforce the incompressibility condition $\nabla\cdot\bm v=0$.

The reversible force density entering Eq.~\eqref{eq:NS} is written as
\begin{equation}
\bm{f}_{\rm rev}(\bm{r}) = -\psi\nabla\mu(\bm{r}) + \sum_{i=1}^{N_\tn p} \frac{\phi_i}{\Omega_i}\bm{f}_i + \bm{f}_{\rm fric}(\bm{r}),
\label{eq:f_rev}
\end{equation}
where $\mu=\delta\mathscr{F}/\delta\psi$ is the chemical potential associated with the composition field, $\bm f_i=-\partial\mathscr{F}/\partial\bm r_i$ is the force acting on particle $i$, and $\Omega_i=\int d\bm r\,\phi_i$ denotes its effective volume. The first term represents capillary forces arising from composition gradients, the second term accounts for particle-induced forces distributed to the fluid through the diffuse-interface representation, and the third term enforces no-slip boundary conditions near the walls. This formulation ensures $d\mathscr{F}/dt\le 0$ while conserving total momentum~\cite{Furukawa2018}.

Equation~\eqref{eq:NS} is solved numerically using the Marker-and-Cell (MAC) method~\cite{Harlow1965,Yuan2022} on a staggered grid with periodic boundary conditions in the lateral directions.

\paragraph*{Composition dynamics.}
The temporal evolution of the composition order parameter $\psi(\bm r,t)$ is governed by a conservation law,
\begin{equation}
\frac{\partial\psi}{\partial t} = -\nabla\cdot\bm{j}_\psi,
\end{equation}
where the flux $\bm j_\psi$ is given by
\begin{equation}
\bm{j}_\psi = \psi\bm{v} - L_\psi\nabla\mu,
\quad L_\psi = L_0\Bigl(1-\sum_i^N \phi_i\Bigr).
\label{eq:flux}
\end{equation}
Here $L_0$ is a constant mobility and $\mu=\delta\mathscr{F}/\delta\psi$ is the chemical potential. The first term in $\bm j_\psi$ represents advective transport of the composition field by the fluid flow, while the second term describes diffusive relaxation driven by chemical potential gradients. The mobility prefactor $L_\psi$ suppresses diffusive transport inside the particle domains, ensuring that compositional diffusion is confined to the surrounding fluid.

\paragraph*{Time integration of particle motion.}
The trajectories of the particles are advanced according to the local fluid velocity field,
\begin{equation}
\frac{d\bm{r}_i}{dt} = \bm{v}_i,
\quad \bm{v}_i = \frac{\int d\bm{r}\,\bm{v}\phi_i}{\Omega_i}.
\label{eq:ri}
\end{equation}
where $\bm v_i$ is the translational velocity of particle $i$ obtained by averaging the fluid velocity over its diffuse interface, weighted by the indicator field $\phi_i$, and $\Omega_i=\int d\bm r\,\phi_i$ is the effective particle volume. This formulation ensures that each particle is advected consistently with the surrounding hydrodynamic flow while remaining smoothly coupled to the fluid through the diffuse-interface representation.

\paragraph{Simulation parameters.}
The lengths are expressed in units of the lattice spacing $\ell_0$, and time is measured in units of $\tau_{s}=\rho\ell_0^{2}/\eta_{\rm s}$. The head and tail radii are set to $a_{\tn h}=a_{\tn t}=3.2\,\ell_0$, and the virtual particle radius to $a_{\tn{vir}}=2.4\,\ell_0$. The diffuse-interface thickness is $\xi=\ell_0$, and the particle viscosity is chosen as $\eta_{\rm p}=50\,\eta_{\rm s}$. The steric repulsion and bond stiffness parameters are $\varepsilon_1=10$ and $\varepsilon_2=3.2\times10^{5}$, respectively, with bond lengths $\ell_1=8.0\,\ell_0$ and $\ell_2=9.6\,\ell_0$. The integration time step is $\Delta t=2.5\times10^{-3}\tau_{\rm s}$.

The Ginzburg--Landau parameters are set to $r_0\epsilon_0=-1$, $u_0=1$, and $k_0=1$. The particle--wall repulsion strength is $\varepsilon=10$, and the wall friction coefficient enforcing the no-slip condition is $\zeta_{\rm w}=200$. The two confining plates are located at $z=0$ and $z=H$, with $H=150\,\ell_0$. Simulations are performed in a cubic domain of lateral size $L=256\,\ell_0$, with periodic boundary conditions applied in the lateral directions.

Active swimmers are modeled with a propulsion force $f_{\tn{act}}=5$, while passive control simulations use $f_{\tn{act}}=0$. In the active case, the estimated Reynolds number is small ($\mathrm{Re}\approx10^{-3}$), confirming that the dynamics occur in the Stokes-flow regime. Additional simulation parameters, including the wetting affinity $\gamma$, the wall wetting parameter $\gamma_{\tn{wall}}$, the average composition $\bar{\psi}$, and the number of bacteria $N_{\tn b}$, are summarized in Table~\ref{tab:fig2_params}. 

All simulations are initialized from a homogeneous composition field with bacteria randomly distributed throughout the simulation box. Computations are performed on an NVIDIA A800 GPU.

\begin{table}[h]
  \centering
  \caption{Parameters corresponding to Fig.~\ref{fig:figure2}a--c and Fig.~\ref{fig:figure6}e.}
  \begin{tabular}{lccc|c}
    \hline
    & Fig.~\ref{fig:figure2}a & Fig.~\ref{fig:figure2}b & Fig.~\ref{fig:figure2}c & Fig.~\ref{fig:figure6}e\\
    \hline
    $\gamma$        & & {$\gamma_1=-4, \gamma_2=2$} & & {$-4$} \\
    \hline
    $\gamma_{\rm wall}$ & & {$-2$} & & {$-2$} \\
    \hline
    $\bar\psi$        & $-1.0$ & $-0.9$ & $-0.6$ & $-0.9$\\
    \hline
    $N_{\rm b}$      & & {$235$} & & $2360$\\
    \hline
  \end{tabular}
  \label{tab:fig2_params}
\end{table}

\subsection{Average number of bacterial neighbors $N_{\mathrm{neigh}}$}
The degree of local bacterial aggregation is quantified through the mean number of neighboring bacteria.
Two bacteria are defined as neighbors if any of the four endpoint separations between their head and tail beads falls below a cutoff distance.
For a pair of bacteria $i$ and $j$, the four separations
$r_{ij}^{\mathrm{HH}}, r_{ij}^{\mathrm{HT}}, r_{ij}^{\mathrm{TH}}, r_{ij}^{\mathrm{TT}}$
are evaluated under periodic boundary conditions (PBCs).

Based on these distances, the adjacency matrix is defined as
\begin{equation}
A_{ij} =
\begin{cases}
1, & \min\!\bigl(r_{ij}^{\mathrm{HH}}, r_{ij}^{\mathrm{HT}}, r_{ij}^{\mathrm{TH}}, r_{ij}^{\mathrm{TT}}\bigr) < r_c,\\
0, & \text{otherwise},
\end{cases}
\qquad A_{ii}=0,
\end{equation}
where the cutoff distance is chosen as $r_c = 4a_{\mathrm{h}} = 4a_{\mathrm{t}}$, determined by the bead size.

The instantaneous number of neighbors of bacterium $i$ is then given by
\begin{equation}
n_i(t)=\sum_{j=1,\, j\neq i}^{N_\tn{b}} A_{ij}(t),
\end{equation}
and the frame-averaged mean neighbor number is defined as
\begin{equation}
N_{\mathrm{neigh}}(t)
=
\frac{1}{N_\tn b}\sum_{i=1}^{N_\tn b} n_i(t).
\end{equation}

\subsection{Accumulated bacterial fraction~$f_{\mathrm{ads}}$}
The population of bacteria accumulated near the confining plates is quantified by the wall-accumulated fraction.
For each bacterium $i$, the distances of its constituent beads
$b\in\{\mathrm{h,t,vir}\}$ from the lower and upper walls are first evaluated as
\begin{equation}
d_{i,n}^{\rm (low)} = z_{i}^{n},
\qquad
d_{i,n}^{\rm (up)}  = H - z_{i}^{n},
\label{eq:nearwall2}
\end{equation}
where $z_i^{n}$ denotes the vertical position of bead $n$.

The minimum wall distance of bacterium $i$ is then defined as
\begin{equation}
d_i
=
\min_{n\in\{\mathrm{h,t,vir}\}}
\bigl\{\min\bigl(d_{i,n}^{\rm (low)}, d_{i,n}^{\rm (up)}\bigr)\bigr\}.
\label{eq:nearwall}
\end{equation}
Bacteria satisfying $d_i<d_{\rm cut}$ are classified as near-wall, with a cutoff distance $d_{\rm cut}=14\,\ell_0$, corresponding approximately to the bacterial length. Throughout the simulation analysis, all references to ``near-wall'' bacteria follow this definition based on the minimum bead--wall distance.

The instantaneous wall-accumulated fraction is then given by
\begin{equation}
f_{\rm ads}(t)
=
\frac{1}{N_{\tn b}}
\sum_{i=1}^{N_{\tn b}}
\Theta\!\bigl(d_{\rm cut}-d_i(t)\bigr),
\end{equation}
where $N_{\tn b}$ is the total number of bacteria and $\Theta$ denotes the Heaviside step function.

\subsection{Near-wall radial distribution function $g(r)$}

To quantify lateral positional correlations among bacteria accumulated near the walls, we compute the radial distribution function (RDF) $g(r)$. The identification of near-wall bacteria follows the same cutoff-based distance criterion used for the accumulated fraction $f_{\rm ads}$. Only frames containing at least two near-wall bacteria ($K\ge2$) are included in the analysis.

For each pair of near-wall bacteria $(i,j)$, the lateral separation is defined as the minimum bead--bead distance projected onto the $xy$ plane,
\begin{equation}
r_{ij}
=
\min_{b,b'}
\left|
\bm r_{i}^{(b)}
-
\bm r_{j}^{(b')}
\right|_{\rm PBC},
\end{equation}
where $b,b'\in\{\mathrm{h,t,vir}\}$ denote the head, tail, and virtual beads, respectively, and periodic boundary conditions (PBCs) are applied in both lateral directions.

These distances are histogrammed into $N_{\mathrm{bin}}=51$ bins over the range $[0,r_{\rm max}]$, where $r_{\rm max}=\min(L_x,L_y)/2$. This yields bin counts $H_\alpha$ and corresponding annular shell areas
\begin{equation}
A_\alpha
=
\pi \left(r_{\alpha+1}^2 - r_\alpha^2\right).
\end{equation}
The areal number density of near-wall bacteria is defined as $\rho=K/(L_xL_y)$.

The RDF in each frame is then computed as
\begin{equation}
g(r_\alpha)
=
\frac{H_\alpha}{\rho\,K\,A_\alpha},
\end{equation}
which quantifies the in-plane spatial correlations of bacteria confined near the walls.

\subsection{Near-wall orientational distribution $P(\theta)$}

The orientation of bacteria accumulated near the confining plates is characterized by the orientational distribution function $P(\theta)$. The set of near-wall bacteria is defined using the same cutoff-based criterion employed for the accumulated fraction $f_{\rm ads}$ and the near-wall RDF analysis.

For each bacterium $i$, the closest plate is identified using Eqs.~(\ref{eq:nearwall2}) and (\ref{eq:nearwall}), which determines the local wall-normal unit vector $\hat{\bm n}_i$ as
\[
\hat{\bm n}_i
=
\begin{cases}
-\hat{\bm z}, & d_i^{\rm (low)} \le d_i^{\rm (up)},\\[2pt]
+\hat{\bm z}, & d_i^{\rm (up)} < d_i^{\rm (low)}.
\end{cases}
\]
The orientation of bacterium $i$ relative to the nearest wall is then quantified by
\begin{equation}
\cos\theta_i = \hat{\bm D}_i\cdot\hat{\bm n}_i
\qquad
\theta_i = \arccos(\cos\theta_i)
\end{equation}
where $\hat{\mathbf{D}_i}=(\bm r_i^{\mathrm{h}}-\bm r_i^{\mathrm{t}})/|\bm r_i^{\mathrm{h}}-\bm r_i^{\mathrm{t}}|$ is is the unit orientation vector of bacterium $i$, pointing from tail to head.

The orientational distribution $P(\theta)$ is obtained by histogramming $\theta_i\in[0,\pi]$ into $N_{\mathrm{bin}}=90$ bins and normalizing such that $\sum_\theta P(\theta)=1$. This distribution characterizes the alignment of bacteria relative to the confining surfaces.

\subsection{Wall-normal momentum flow}

Transport of momentum toward the confining plates is quantified by the wall-normal momentum flow,  $\langle \bm v_z \cdot \bm n \rangle(t)$.
For each particle $i$, the unit normal vector $\bm n_i$ is defined to point toward the nearest wall,
$$\bm n_i
=
\begin{cases}
-\hat{\bm z}, & z_i < H/2,\\[2pt]
+\hat{\bm z}, & z_i \ge H/2 ,
\end{cases}
$$
thereby specifying the direction of adsorption onto the closest plate.

At each time point, the spatially averaged wall-normal velocity is computed as
\begin{equation}
\langle \bm v_z \cdot \bm n \rangle(t)
=
\frac{1}{N_\tn p}
\sum_{i=1}^{N_\tn p}
\bm v_{z, i}\cdot\bm n_i,
\end{equation}
where $\bm v_{z,i}$ is the $z$ component of the velocity of particle $i$, and the sum runs over all particles ($N_{\tn p}$ in total). A positive value of $\langle \bm v_z \cdot \bm n \rangle$ indicates a net momentum flux directed from the bulk toward the confining walls, whereas a negative value corresponds to momentum transport away from the interfaces.

\subsection{Mean bacterium-wall distance $d_{\rm wall}(t)$}
The proximity of bacteria to the confining plates is quantified by the mean minimum bacterium--wall distance. For each bacterium $i$, the instantaneous minimum distance to either wall is defined as
\begin{equation}
d_i^{\rm wall}(t)
=
\min_{b\in\{{\rm h,t,vir}\}}
\Bigl\{ z_i^{(b)}(t),\,H-z_i^{(b)}(t) \Bigr\},
\end{equation}
where $z_i^{(b)}(t)$ denotes the vertical position of bead $b$ of bacterium $i$ at time $t$.

The mean distance to the wall is then obtained by averaging over all bacteria,
\begin{equation}
d^{\rm wall}(t)
=
\frac{1}{N_\tn b}
\sum_{i=1}^{N_\tn b} d_i^{\rm wall}(t),
\end{equation}
which provides a time-resolved measure of the overall confinement of bacteria near the interfaces.

\begin{figure}[t!]
\centering \includegraphics[width=8.5cm]{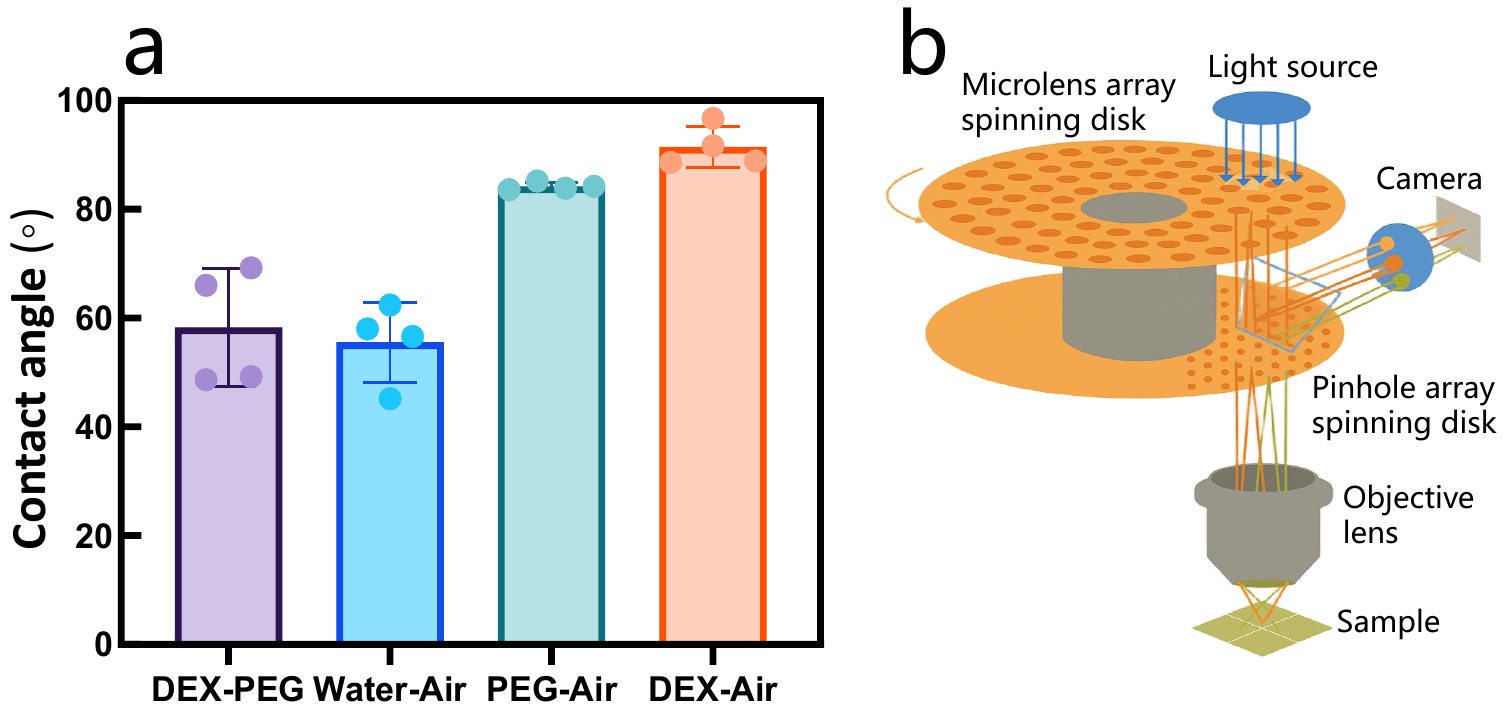}
\caption{Wettability characterization of \textit{P. aeruginosa} and experimental imaging setup.
(a) Contact angle measurements performed on \textit{P. aeruginosa} bacterial lawns. Shown are the contact angles of the DEX-rich phase in a PEG-rich background (DEX--PEG), as well as those of water, the PEG-rich phase, and the DEX-rich phase at the air--bacteria interface. The DEX--PEG contact angle ($<90^{\circ}$) indicates a preferential affinity of the bacterial surface for the DEX-rich phase~\cite{yang2025active}.
(b) Schematic illustration of the spinning-disk confocal microscopy setup used for imaging. A microlens- and pinhole-array spinning disk enables high-speed optical sectioning for three-dimensional visualization of bacterial distributions.}
\label{fig:figureS1}
\end{figure}

\begin{figure}[t!]
\centering \includegraphics[width=8.5cm]{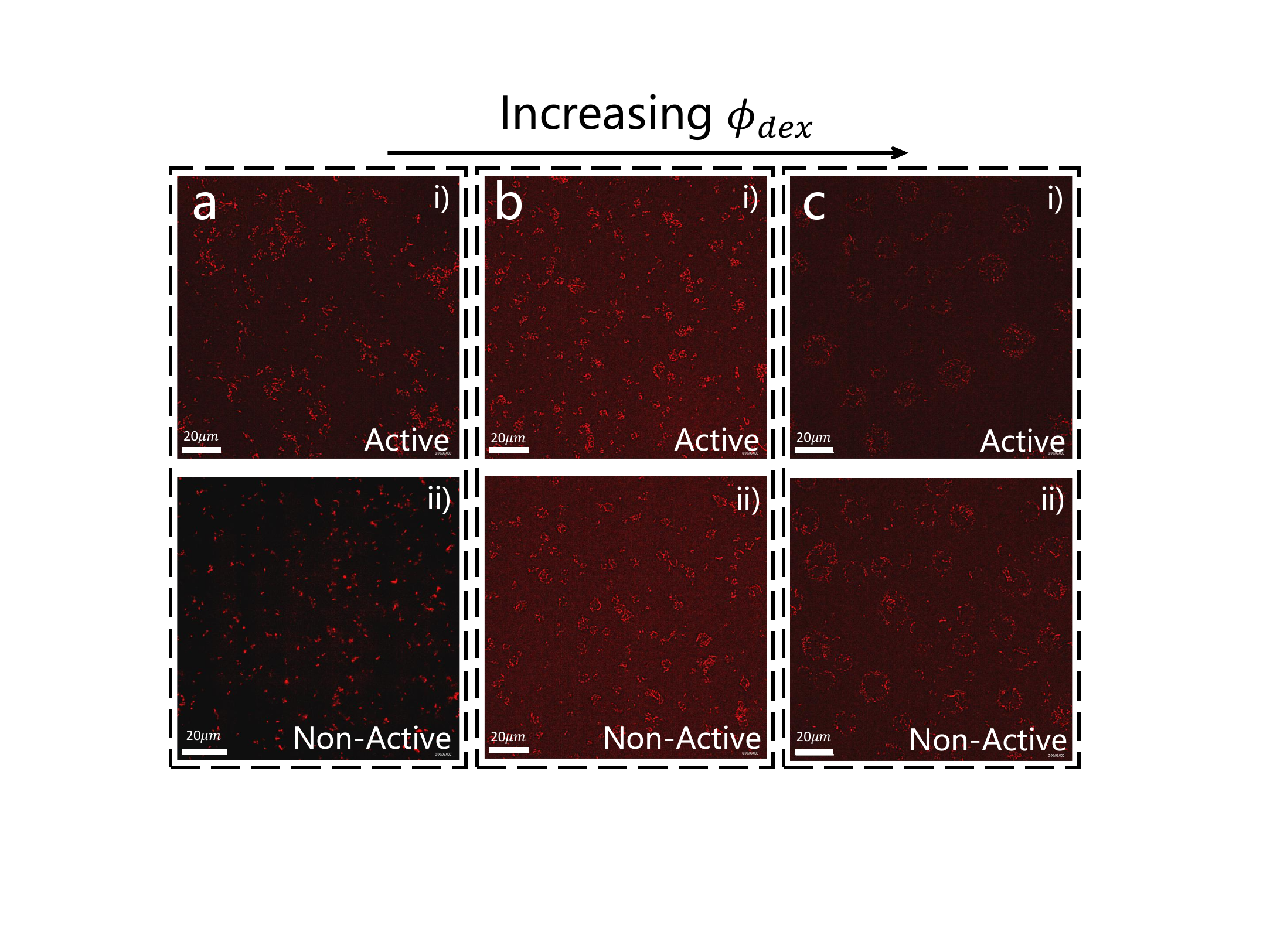}
\caption{Comparison of interfacial accumulation between active and non-active bacteria across different phase regimes.
(a--c) Representative confocal microscopy images of active (top row, i) and non-active (bottom row, ii) \textit{P. aeruginosa} (red) at increasing volume fractions of the DEX-rich phase. Bacteria within a 7.5~$\mu$m-thick layer adjacent to the interface are visualized.
(a) At $\phi_{\tn{dex}}=0$, both active and non-active bacteria exhibit minimal interfacial accumulation.
(b) At a low phase fraction $\phi_{\tn{dex}}=0.05$, the formation of a surface-associated wetting layer leads to pronounced accumulation for both cases, with enhanced coverage relative to (a).
(c) At high phase fraction $\phi_{\tn{dex}}=0.2$, the minority phase forms droplets. Non-active bacteria (ii) are passively carried by these droplets and settle at the interface, producing visible ring-like structures. In contrast, active bacteria (i) exhibit strongly suppressed accumulation, appearing as a darker field, consistent with motility-driven inhibition of settling due to self-spinning droplets as discussed in the main text.}
\label{fig:figureS2}
\end{figure}

\begin{figure}[t!]
\centering \includegraphics[width=8.5cm]{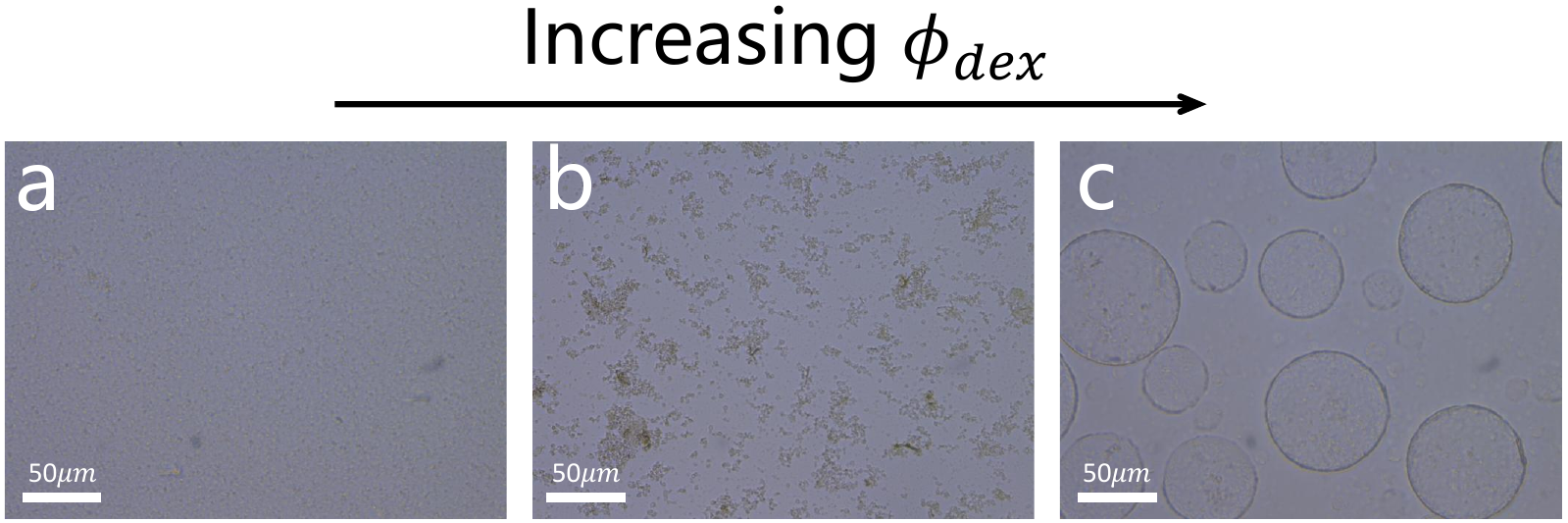}
\caption{Bright-field images showing the accumulation of active \textit{P. aeruginosa} under different phase conditions.
(a) Minimal interfacial accumulation in the absence of DEX.
(b) Enhanced accumulation mediated by the formation of a surface-associated wetting layer at $\phi_{\tn{dex}}=0.05$.
(c) Confinement of bacteria within DEX-rich droplets at $\phi_{\tn{dex}}=0.2$.}
\label{fig:figureS3}
\end{figure}

\begin{figure}[t!]
\centering \includegraphics[width=8.5cm]{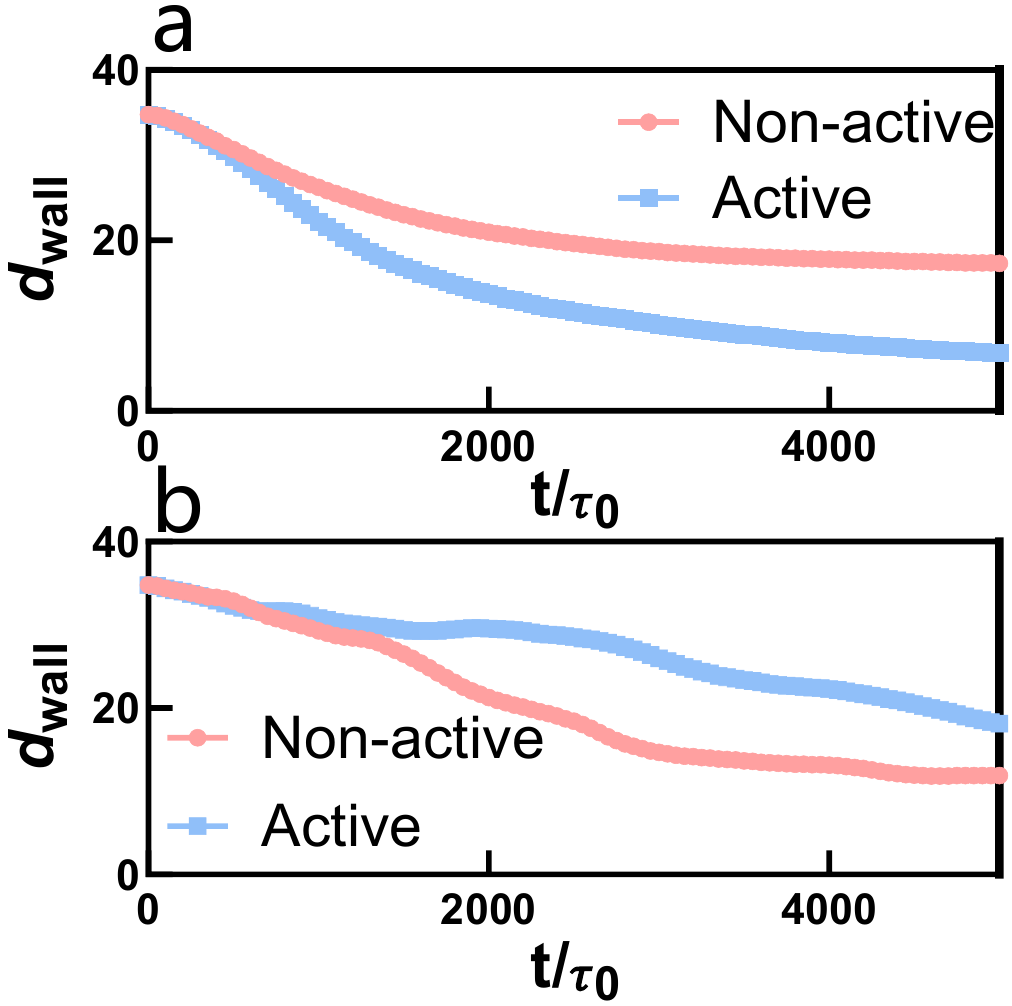}
\caption{Quantification of bacterial adhesion.
(a) Simulated mean bacterium--wall distance $d_{\text{wall}}$ for $\phi_{\tn{dex}}=0.05$ corresponding to Fig.~\ref{fig:figure3}. Active bacteria exhibit a smaller $d_{\text{wall}}$ than non-active bacteria, indicating enhanced interfacial accumulation.
(b) Mean bacterium--wall distance $d_{\text{wall}}$ for $\phi_{\tn{dex}}=0.2$ corresponding to Fig.~\ref{fig:figure4}. In this case, $d_{\text{wall}}$ is larger for active bacteria, consistent with motility-induced suppression of interfacial accumulation.}
\label{fig:figureS4}
\end{figure}

\begin{figure}[t!]
\centering \includegraphics[width=8.5cm]{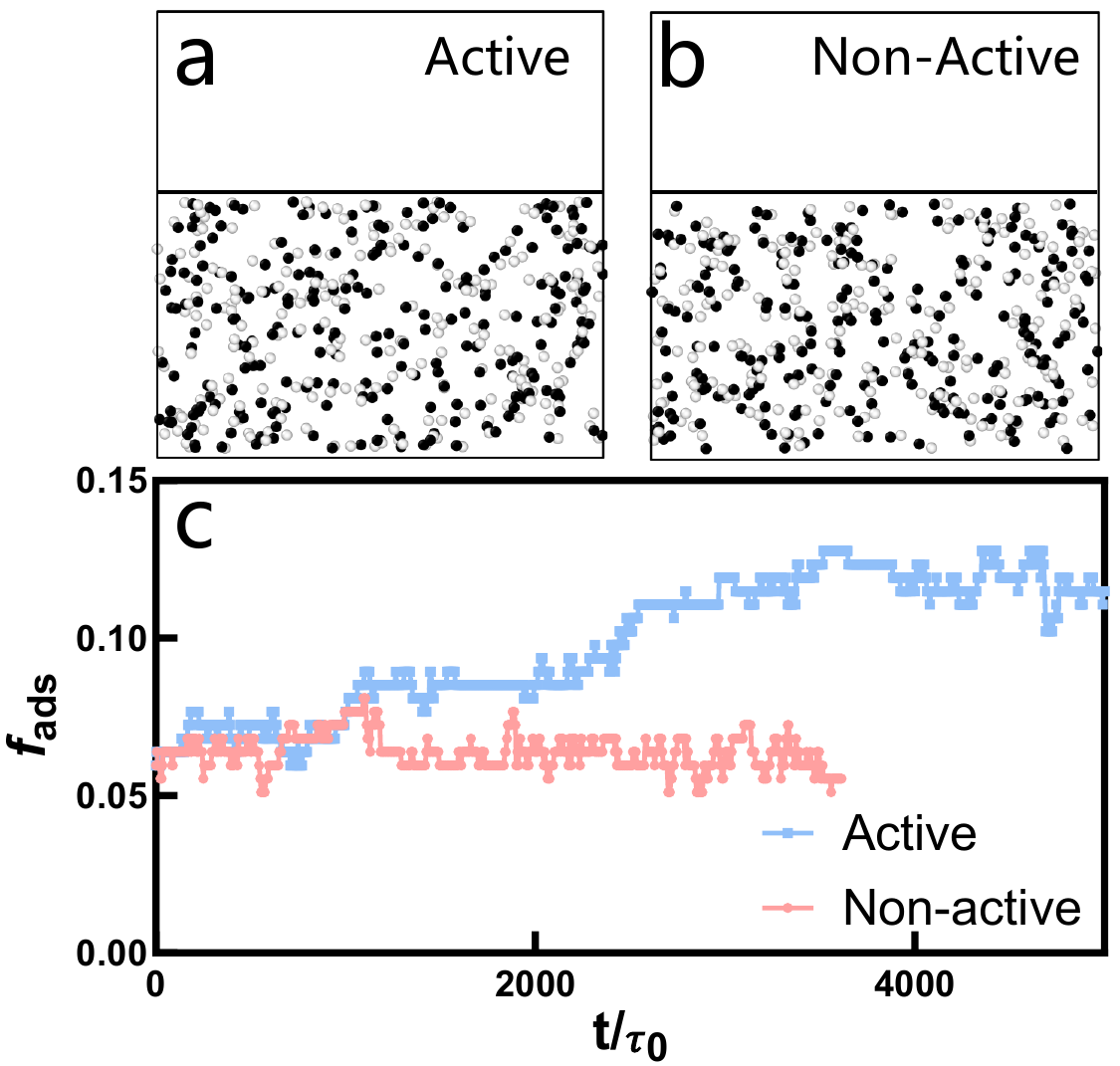}
\caption{Simulated bacterial accumulation in a homogeneous single-phase control.
Temporal evolution of the accumulated bacterial fraction $f_{\tn{ads}}$ obtained from simulations in a homogeneous solution ($\phi_{\tn{dex}}=0$). In the absence of a DEX-rich wetting layer, interfacial accumulation remains minimal ($f_{\tn{ads}}<0.15$) for both active and non-active bacteria. This behavior contrasts sharply with the phase-separated systems, demonstrating that the pronounced accumulation reported in the main text is driven by the wetting-mediated mechanism rather than by intrinsic bacteria--wall attraction or hydrodynamic trapping.}
\label{fig:figureS5}
\end{figure}

\begin{figure}[h!]
\centering \includegraphics[width=8.5cm]{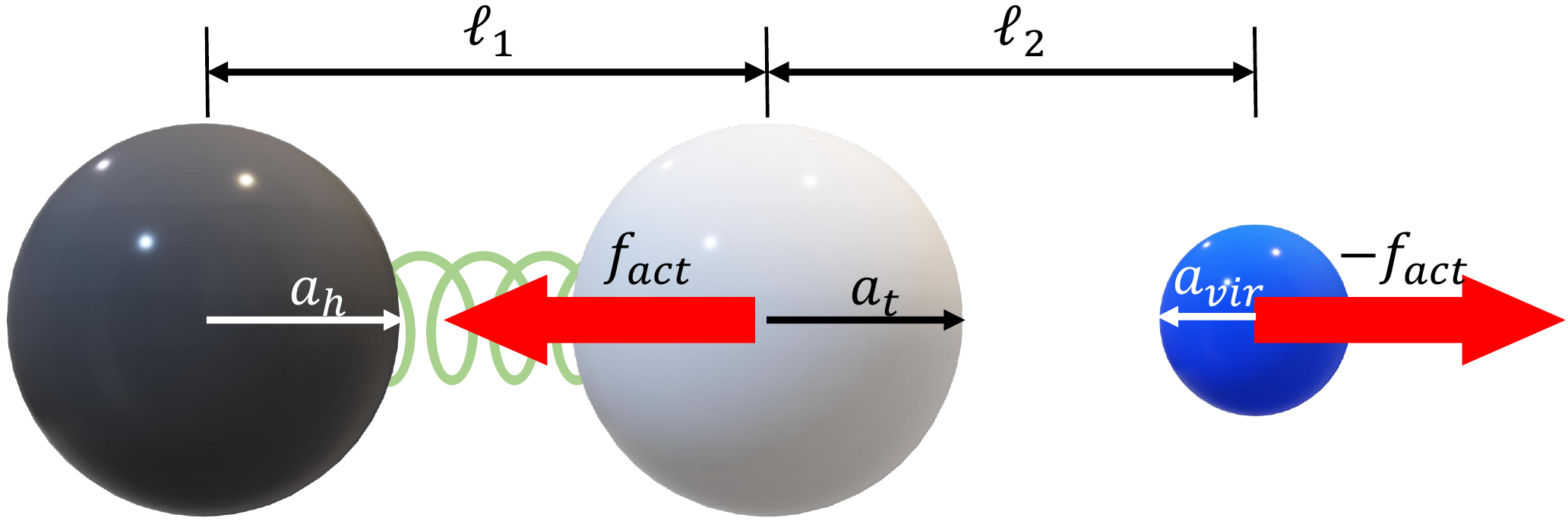}
\caption{Schematic illustration of the coarse-grained bacterial model used in simulations~\cite{furukawa2014activity}. The bacterial model consists of a head (black bead), an active tail (white bead), and a virtual particle (blue bead). The head and tail beads are connected by a harmonic spring with bond length $\ell_1$. Red arrows indicate the equal and opposite active forces, $\pm f_{act}$, applied to the tail and virtual particles. The active forces drive the system out of equilibrium while ensuring momentum conservation without external drift. $\ell_2$ denotes the distance between the tail and the virtual particle.}
\label{fig:figureS6}
\end{figure}

\FloatBarrier

%

\end{document}